\documentclass[namedreferences]{solarphysics}

\usepackage[hyperref,optionalrh,showbiblabels]{spr-sola-addons} 
\usepackage{graphicx}        
\usepackage{amssymb}        
\usepackage{color}           
\usepackage{amsfonts}
\usepackage[normalem]{ulem} 
\usepackage{rotating}
\usepackage[graphicx]{realboxes}
\usepackage{pdflscape}
\usepackage{afterpage}
\usepackage{threeparttable}
\usepackage{adjustbox}
\hypersetup{
    colorlinks=true,
    linkcolor=blue,
    citecolor=blue
}

\renewcommand{\vec}[1]{{\mathbfit #1}}

\newcommand{\aap}{    {\it Astron. Astrophys.}}
\newcommand{\aaps}{   {\it Astron. Astrophys. Suppl.}}

\newcommand{\apj}{    {\it Astrophys. J.}}
\newcommand{\apjl}{   {\it Astrophys. J. Lett.}}

\newcommand{\nat}{    {\it Nature}}

\newcommand{\pasj}{   {\it Pub. Astron. Soc. Japan}}

\newcommand{\solphys}{{\it Solar Phys.}}
 
\newcommand{\ssr}{    {\it Space Sci. Rev.}}
\newcommand{\apjs}{   {\it Astrophys. J. Suppl. Series}}

\begin{document}

\begin{article}
\begin{opening}

\title{Spectral lines in FUV and EUV for diagnosing coronal magnetic field}

\author[addressref={aff1},corref,email={raveena.khan@iiap.res.in}]{\inits{Raveena}\fnm{Raveena}~\lnm{Khan}\orcid{0000-0001-6104-8938}}
\author[addressref=aff1,corref,email={nagarajuk@iiap.res.in}]{\inits{K.}\fnm{K.}~\lnm{Nagaraju}\orcid{0000-0002-0465-8032}}

\address[id=aff1]{Indian Institute of Astrophysics, Bengaluru - 560034, India}

\runningauthor{R. Khan and K. Nagaraju}
\runningtitle{UV spectral lines for diagnosing coronal magnetic field}

\begin{abstract}
The diagnostic capabilities of spectral lines in far ultraviolet (FUV) and extreme ultraviolet (EUV) wavelength range are explored in terms of their Hanle and Zeeman sensitivity to probe vector magnetic field in the solar corona. The temperature range covered is log$_{10}(T)=5.5-6.3$.  The circular polarization signal due to longitudinal Zeeman effect is estimated for spectral lines in the wavelength range of 500 to 1600 \AA. The Stokes~$V/I$ signal for a FUV line is found to be in the order of 10$^{-4}$ for a longitudinal field strength of 10~Gauss, which further reduces to 10$^{-5}$ for wavelengths below 1200 \AA. 
Due to such low signals, the present study aims to find  combination of spectral lines having different Hanle sensitivity but with identical peak formation temperature to probe coronal magnetic field vector. The combination of Hanle sensitive lines is better suited  because the Hanle signals are stronger by at least an order of magnitude compared to Zeeman signals. 
The linear polarization signals due to Hanle effect from at least two spectral lines are required to derive information on the full vector. It is found from this study that there is always a pair of Hanle sensitive lines for a given temperature range suitable for probing coronal vector magnetic field and they are located in close proximity with each other in terms of their wavelength.
 
\end{abstract}

\keywords{Solar corona, magnetic field, Zeeman effect $\cdot$ Ultraviolet $\cdot$ Polarization, Hanle effect}
\end{opening}

\section{Introduction}
     \label{Sec:Intro} 
The dominance of the magnetic field in the complex structuring of the solar corona is on account of the low plasma $\beta$ (the ratio of kinetic pressure to magnetic pressure). There is a rapid decrease in plasma $\beta$ while going from photosphere to corona due to which the equilibrium field becomes force-free in the corona where magnetic pressure dominates over gas pressure \citep{Priest1991gamp.conf.....P}. 
These magnetic fields play a fundamental role in the formation and evolution of coronal features like coronal loops and streamers, in governing thermal and magnetohydrodynamic (MHD) characteristics of the corona, which in turn regulate the phenomena such as plasma heating, particle acceleration and explosive events.
The coronal magnetic fields drive most of the solar events such as flares, jets, coronal mass ejections (CMEs) and solar energetic particles. 
Information on magnetic field vector and its dynamic evolution is required in the solar atmosphere, more crucially in corona, to understand the exact role it plays in driving these events. 

Routine magnetic field measurements are being carried out in the photosphere with high spatial resolution over a selected region of interest as well as over full disk with moderate resolution  \citep{Lagg2017SSRv..210...37L}. 
In the recent past, significant progress has been made with magnetic field measurements in the chromosphere as well \citep{Trujillo2014ASPC..489..137T, Lagg2017SSRv..210...37L, Ishikawa2021SciA....7.8406I}. On the other hand, coronal magnetic field measurements are still sporadic. Confirmed detection of Stokes~$V$ signal through Zeeman effect in the forbidden line due to Fe {\sc xiii} at 10747 \AA~had been reported by \cite{Lin2000ApJ...541L..83L} and \cite{Lin2004ApJ...613L.177L}. \cite{RaouafiOVIdetect1999A&A...345..999R} reported the linear polarization signal in O {\sc vi} at 1032 \AA~from the Solar and Heliospheric Observatory (SOHO)/Solar Ultraviolet Measurements of Emitted Radiation (SUMER) spectroscopic observations that was recorded during the roll manoeuver of the SOHO satellite. \cite{RaouafiOVIdetect1999A&A...345..999R, RaouafiOVImeasurepol2002A&A...390..691R, RaouafiOVIBmeasure2002A&A...396.1019R} interpreted this signal in terms of Hanle effect and derived a field strength of $\approx 3 $~G at 0.3 $R_\odot$ above a coronal hole. The Coronal Multi-channel Polarimeter (CoMP) produced full Stokes spectropolarimetric measurements in coronal emission lines due to Fe {\sc xiii} at 10747~\AA ~and 10798~\AA ~and, chromospheric line He {\sc i} at 10830~\AA~\citep{Tomczyk2008SoPh..247..411T}. In spite of full Stokes spectropolarimetry, the measurements had been mainly used to study magnetic topology in line-of-sight (LOS) aligned structures such as, pseudostreamers \citep{Gibson2017ApJ...840L..13G} and coronal cavities \citep{BS2013ApJ...770L..28B} and, to detect waves for deducing transverse component of the magnetic field \citep{McIntosh2011Natur.475..477M,YangGlobal2020Sci...369..694Y}. In near future, high resolution and high precision spectropolarimetric observations from Daniel K. Inouye Solar Telescope (DKIST) may become available for coronal magnetometry, over the wavelength range of 3800 to 50000 \AA ~\citep{Rast2021SoPh..296...70R}. ADITYA-L1, a space based observatory which is expected to be launched in near future, is also expected to produce spectropolarimetric observations of corona in  Fe {\sc xiii} 10747~\AA~line \citep{RaghavendraPrasad2017CSci..113..613R, Nagaraju2021ApOpt..60.8145N}. Another upcoming ground-based facility is the COronal Solar Magnetism Observatory (COSMO) which will comprise of the Large Coronagraph (LC), the K-coronagraph (K-cor) and the Chromospheric and Prominence Magnetometer (ChroMag) for the measurement of magnetic fields and thermodynamic conditions in the chromosphere and corona \citep{Tomczyk2016JGRA..121.7470T}. Apart from observations in ultraviolet/visible/infrared wavelengths, magnetic field measurements in radio and microwave wavelengths have also been reported \citep{Peter2012ExA....33..271P, Kishore2015SoPh..290.2409K, Mugundhan2018SoPh..293...41M, Kumari2019ApJ...881...24K}. A novel spectroscopic technique, so called Magnetic-field induced transition (MIT), is being employed recently to infer magnetic field strength in the corona \citep{Li2016ApJ...826..219L, Landi2020ApJ...904...87L, Landi2021ApJ...913....1L, Li2021ApJ...913..135L}.

The comprehension of large number of physical processes taking place in corona requires accurate knowledge about vector magnetic field simultaneously at multiple heights. The coronal field measurements reported above have one or more limitations to infer the vector magnetic field in the corona. The linear polarization signal of forbidden lines is practically insensitive to magnetic field strength, and it can only constrain the field orientation in the plane perpendicular to LOS (field azimuth) \citep{Casini1999ApJ...522..524C} through the saturated Hanle effect. On the other hand, for permitted lines which fall within the regime of the unsaturated Hanle effect, the linear polarization is in theory sensitive to the full vector magnetic field. With both permitted and forbidden coronal emission lines, the LOS component of the field produces circular polarization through longitudinal Zeeman effect; however, the circularly polarized signal induced by the coronal magnetic field is very weak. The measurements by \cite{Lin2004ApJ...613L.177L} have shown that the Stokes~$V/I$ signal in Fe~{\sc xiii} at 10747 \AA~is close to $10^{-4}$ for a LOS field strength of a few Gauss. This signal is expected to be about an order of magnitude smaller in far ultraviolet (FUV) wavelengths for a line with comparable effective Land\'e factor and line steepness (i.e. $\frac{dI}{d\lambda}$). Owing to wavelength  scaling of the Stokes~$V/I$ signal, its amplitude will be even smaller at extreme ultraviolet (EUV) wavelengths. Given the difficulties in measuring such low signals, the unsaturated Hanle effect offers a clear advantage as full vector field diagnostic. \cite{Bommier1981A&A...100..231B} proposed a method which utilises a minimum of two permitted lines with different Hanle sensitivity to obtain the vector magnetic field information. This method has successfully been applied to derive vector magnetic field in the prominences \citep{Bommier1994SoPh..154..231B, Bommier2021A&A...647A..60B}, but can in principle be extended to the study of coronal fields. Spectral lines in UV (FUV and EUV) may be best suited to infer the coronal magnetic field vector since, there are several permitted lines located in these wavelength ranges which exhibit varied Hanle sensitivity.

Another advantage of UV lines is that both off-limb as well as on-disk measurements can be carried out at these shorter wavelengths, whereas only off-limb observations are possible in visible/infrared (vis/IR) wavelengths. On-disk measurements are best suited for deriving magnetic field stratification in the solar atmosphere.
An amalgamation of UV lines may be used to infer magnetic field at multiple heights since, there are spectral lines in FUV and EUV which form at different heights all the way from photosphere to corona through chromosphere and transition region. One may also choose to combine on-disk photospheric and chromospheric observations in vis/IR with coronal measurements in UV.

There are several studies dedicated to exploiting the capability of \textit{individual} spectral lines in FUV and EUV to probe the coronal magnetic field. A chief FUV line that has been extensively studied is the O~{\sc vi} at 1032~\AA~\citep{SahalOVI1986A&A...168..284S, RaouafiOVImeasurepol2002A&A...390..691R, RaouafiOVIBmeasure2002A&A...396.1019R, Trujillo2017SSRv..210..183T, Zhao2019ApJ...883...55Z}. Another FUV line which has been even more extensively studied is the Ly-$\alpha$ at 1216~\AA~ \citep{bommier1982SoPh...78..157B, Trujillo2014ASPC..489..137T, Kano2017ApJ...839L..10K, Kano2019AAS...23430216K, Hebbur2021ApJ...920..140H}. Nevertheless, multiple line diagnostics will provide redundancy which, of course, will help in overcoming the uncertainties and ambiguities in the vector field measurement. There are only a few papers dedicated to such studies. For example, \cite{Brechot1981SSRv...29..391S} has given a table of spectral lines consisting of both FUV lines and low temperature IR lines which form either in upper chromosphere or transition region. \cite{Judge1998ApJ...500.1009J} has provided a list of forbidden lines in the infrared which are potential diagnostics for probing coronal magnetic field. The current paper is about searching for spectral lines and their combination in FUV and EUV to probe vector magnetic field in the solar corona. The line combinations are chosen such that their formation temperatures are comparable since, it is most likely that they would be originating from the same coronal features. 

\section{FUV and EUV Spectral Lines and their magnetic sensitivity}
Diagnostic capabilities of spectral lines in FUV and EUV in terms of Hanle and Zeeman effects are quantified in this section.
Essence of Hanle effect in diagnosing the magnetic field is the modification of scattering polarization (i.e., linear polarization) in spectral lines and rotation of plane of polarization in the presence of external magnetic field. Such an effect is observed when the splitting of energy levels of a given spectral line due to external magnetic field is comparable to their natural broadening. This implies that the Hanle effect is most effective when \citep{Bommier1981A&A...100..231B}
\begin{equation}
    \label{eq:HE_cond1}
    g_{u} \omega_B \tau = 1,
\end{equation}
where $g_{u}$ is the Land\'e factor of the upper atomic level; $\omega_B$ is the Larmor frequency and $\tau$ is the lifetime of the upper energy level, which is equivalent to the reciprocal of summation over the Einstein $A$ coefficients, assuming that radiation and collision induced transitions are negligible with respect to the spontaneous radiative de-excitation.

The Larmor frequency for a given magnetic field strength $B$ is given by
\begin{equation}
    \label{eq:Lm_freq}
    \omega_{B}=\frac{\mu_{B}}{\hbar}B,
\end{equation} 
where $\mu_{B}$ is Bohr magneton; and $\hbar$ is the reduced Planck constant. When eq. \ref{eq:HE_cond1} is satisfied, the corresponding field strength is called the critical field (B$_{H}$). \cite{Bommier1981A&A...100..231B} have defined the domain of Hanle sensitivity as 
\begin{equation}
    0.1\le g_{u} \omega_{B}\tau \le 10
        \label{eq:HE_cond2}
\end{equation}
based on the uncertainty analysis of vector field determination. In the lower limit, the relative error on field strength is small but uncertainty in determining the field direction is large. In the upper limit, it is the vice-versa. The condition \ref{eq:HE_cond2} is further restricted to the domain \citep{Trujillo2014ASPC..489..137T}
\begin{equation}
    0.1\le g_{u} \omega_{B}\tau \le 5
        \label{eq:HE_cond3}
\end{equation}
which is used in the current work for the selection of suitable spectral lines with varied Hanle sensitivity.

Zeeman effect is caused by the splitting of the energy levels in the presence of external magnetic field which produces characteristic polarization depending on the orientation of vector magnetic field with respect to the observer's LOS. Given the expected field strength in the corona, permitted lines are mostly in the Hanle regime due to their shorter lifetimes (in the order of 10$^{-8}$~s). On the other hand, magnetic sub-levels of the forbidden lines are well separated (i.e., $g_{u} \omega_{B}\tau \gg 5$, which breaks the condition \ref{eq:HE_cond3}). As a consequence, only the LOS field strength through circular polarization \citep{Harvey1969PhDT.........3H} and the field azimuth through linear polarization \citep{Querfeld1984SoPh...91..299Q,Arnaud1987A&A...178..263A} can be determined from these lines. This is because the linear polarization produced by the transverse Zeeman effect is below the detection level of current observational capabilities for magnetic fields of a few Gauss. Thus, Stokes $Q$ and $U$ are completely dominated by the residual atomic alignment\footnote{Atomic alignment is defined as the differential pumping of the atomic states, with magnetic quantum numbers $|M|$, due to anisotropic illumination of the atoms.} which does not depend on the field strength (so-called saturated Hanle effect: \footnote{In the saturation limit of the Hanle effect, all the quantum level coherence is destroyed, while only the population imbalance due to the anisotropic radiation remains, which is insensitive to the field strength.} \cite{Sahal1977ApJ...213..887S}). The circular polarization (i.e. Stokes $V$) is also weak but a few measurements had been reported in the literature. For example, the observations carried out by \cite{Lin2000ApJ...541L..83L} and \cite{Lin2004ApJ...613L.177L} had shown the Stokes $V$ amplitudes in the order of $10^{-4}$ for the longitudinal field strength of a few Gauss. The observations were carried out in Fe~{\sc xiii} line at 10747 \AA. For the emission lines in FUV and EUV, this signal is expected to be weaker by at least an order of magnitude because of the way the circular polarization sensitivity index ($s_V$) scales with the wavelength given as (same as Eq. 9.88 of \cite{Landi2004ASSL..307.....L}, but calculated for an emission line with Gaussian profile)
\begin{equation}
    s_V = \left( \frac{\lambda_0}{\lambda_\mathrm{ref}} \right) \overline{g} d_c,
        \label{eq:StokesVsensitivity}
\end{equation}
where $\lambda_0$ is the central wavelength of the emission line under consideration; $\lambda_\mathrm{ref}$ is a reference wavelength (e.g., 1242 \AA);  $\overline{g}$ is the effective Land\'e factor;  and $d_c= 1 - I(\lambda_0)/I_c$, is the line centre depression (negative) with $I_c$ being the continuum intensity adjacent to the spectral emission line. The basis for deriving eq.~\ref{eq:StokesVsensitivity} is the equation that relates Stokes $V$ to the longitudinal magnetic field (B$_{\parallel}$) under weak field approximation \citep{Landi2004ASSL..307.....L} which is given by
\begin{equation}
     \frac{V (\lambda)}{I(\lambda)}= -4.67\times10^{-13}~ \overline{g} \lambda^{2}B_{\parallel} \frac{1}{I(\lambda)} \frac{dI(\lambda)}{d\lambda}.
     \label{eq:Vsignal}
\end{equation}
In the above equation  $\frac{dI(\lambda)}{d\lambda}$ is the intensity derivative with respect to wavelength. As can be seen from eq.~\ref{eq:Vsignal}, apart from the wavelength dependence, Stokes $V$ amplitude also depends on $\overline{g}$ and  $\frac{dI(\lambda)}{d\lambda}$ which usually differ from one spectral line to the other. Hence, it is worth checking for the Stokes $V$ sensitivity of the spectral lines in UV so that this information may be combined with the Hanle measurements to infer the vector magnetic field. 

As can be seen from eqs. \ref{eq:HE_cond1} to \ref{eq:Vsignal}, effective Land\'e factor, Einstein $A$ coefficients and $\frac{dI(\lambda)}{d\lambda}$ corresponding to a given spectral line are some of the essential parameters to determine its magnetic sensitivity. In addition, as spectropolarimetric measurements are in general photon starved, line irradiance is also taken into account while checking for the diagnostic potential of a given spectral line to probe the magnetic field. The estimation or compilation of these parameters is discussed briefly in the following subsections. 

\subsection{Spectral lines}
In this analysis, spectral lines observed from three different space-based missions are compiled. One is the EUV Imaging Spectrometer (EIS; \citealp{Culhane2007ASPC..369....3C}), onboard the Hinode spacecraft, which is designed to observe in two wavelength ranges (SW: 166 – 212 \AA; LW: 245 – 291 \AA). These wavelength bands consist of several emission lines from highly ionised species ranging from 4.7 to 7.3 in the logarithmic scale of temperature (log$_{10}$(T)). Another is a sounding rocket instrument, named the Extreme Ultraviolet Normal Incidence Spectrograph (EUNIS), which observed a coronal bright point around 18:12 UT on 2006 April 12. A brief description of the instrument has been provided by \cite{Brosius2007ApJ...656L..41B}. The EUV spectra obtained by EUNIS covers first-order wavelengths from 300 to 370 \AA ~over a temperature range of 5.2 to 6.4 in log$_{10}$(T) \citep{Brosius2008ApJ...677..781B}. The third instrument is SUMER which is a high-resolution telescope and spectrograph, onboard the Solar and Heliospheric Observatory (SOHO), which observed the sun over the wavelength range from 470 to 1609~\AA~\citep{curdt2001spectral, curdt2004sumer}.

The spectral lines thus compiled are listed in Table~\ref{Tab:eff_g}.  The atomic species along with their ionization state are listed in the first column and the observed wavelength in Angstrom (\AA)~is given in the second column. In the third column listed are the corresponding level configurations. The fourth and the fifth columns list the peak formation temperature (logarithmic value) of the lines and the transition type, respectively. These line formation temperatures are taken from \cite{feldman1997, Brosius2008ApJ...677..781B, curdt2001spectral, del2005spectral, young2007euv, moran2003}. Although the peak formation temperature is mentioned for most of the spectral lines in Table~\ref{Tab:eff_g}, it should be noted that the lines are always formed in a range of temperature, not at a single value \citep{Feldman1998ApJ...502..997F, Warren2002ApJ...571..999W, Warren2009ApJ...700..762W, Saqri2020SoPh..295....6S}. For example, any spectral line shown at $\mathrm{log}_{10}(T)=6.1$ may actually be formed within a range of $\mathrm{log}_{10}(T)=5.85$ to 6.3. Further, the critical field strength (in Gauss) for maximum Hanle sensitivity are listed in the ninth column. The remaining columns are described in the following subsections. 

\subsection{Transition probability}
The lifetime of the upper energy level is equal to the reciprocal of sum over the Einstein $A$ coefficients, i.e. $\tau = \frac{1}{\sum_j A_{kj}}$. Transition probability or the Einstein $A$ coefficient ($A = \sum_j$ A$_{kj}$) is the total rate of all spontaneous radiative transitions from the upper level (\textit{k}) to all the lower levels to which the level \textit{k} can de-excite. These coefficients help in determining the range of magnetic field strength to which the spectral lines are sensitive in the Hanle regime (cf. eqs.~\ref{eq:HE_cond1} and \ref{eq:HE_cond2}). Besides, they are also important for the visibility of the spectral lines in the solar corona. The $A$ values are obtained from current version 10.0 of the CHIANTI atomic database \citep{Dere1997A&AS..125..149D, DelZanna2021ApJ...909...38D} and are listed in the sixth column of Table~\ref{Tab:eff_g}.

\subsection{Land\'e factors}
In light elements, the electrostatic interaction dominates over the spin-orbit coupling such that the orbital angular momenta of the individual electrons get coupled to give a total orbital angular momentum $\vec{L}$, while the spins of the electrons get coupled to give a total spin $\vec S$. This is referred to as Russell-Saunders or LS coupling. However, for heavier atoms with larger nuclear charge, the spin-orbit interactions are stronger leading to jj coupling. A more general and practical case exists in certain atoms, particularly mid-weight atoms and those with almost closed shells, which lie in between these two coupling limits. Such a coupling is termed as intermediate or i-coupling, in which both the electrostatic and the spin-orbit interactions may be present with a relative order of magnitude. From the selection rules described in \cite{Condon1935tas..book.....C} and \cite{Drake2006sham.book.....D}, the coupling scheme of the atomic levels associated with the dipole (electric or magnetic) transitions are identified. In case of LS coupling, the following expression is used to determine the Land\'e factors of individual energy levels.
\begin{equation}
    g = \frac{3}{2} + \frac{S( S+1) -  L( L+1)}{2 J( J+1)}
    \label{eq:g_LS}
\end{equation}
where $ L$ and $ S$ are the total orbital and spin angular momentum quantum numbers, respectively; and $J$ is the total angular momentum which is defined as $\vec J = \vec L + \vec S$. Equation \ref{eq:g_LS} holds only for LS coupling scheme which may fail in cases involving atomic or ionic lines of high excitation potential and intermediate coupling may have to be considered. There are no lines exhibiting jj coupling in this analysis. For intermediate coupling, the Land\'e factors $g_{1}$ and $g_{2}$ are first estimated using formula \ref{eq:g_LS} and then compared with the values (for those available) from \cite{verdebout2014}. It is found that the values calculated in this work match closely with those estimated using the GRASP2K \citep{Jonsson2007CoPhC.177..597J} code in \cite{verdebout2014}. GRASP2K is a fully relativistic multiconfiguration Dirac-Hartree-Fock (MCDHF) method based atomic structure package. The Land\'e factors of the individual levels are listed in the seventh column of Table~\ref{Tab:eff_g}.

The effective Land\'e factors are calculated using the following formula \citep{LandiDeglInnocenti1982SoPh...77..285L}.
\begin{equation}
    \overline{g}=\frac{1}{2}(g_{1}+g_{2}) +\frac{1}{4}(g_{1}-g_{2})[ J_{1}( J_{1}+1)- J_{2}( J_{2}+1)]
    \label{eq:g}
\end{equation}
where $ J_{1}$ and $g_{1}$ are the total angular momentum and Land\'e factor of the lower energy level, respectively; $ J_{2}$ and $g_{2}$ correspond to that of the upper energy level. All the estimated values of $\overline{g}$ factor are listed in the eighth column of Table~\ref{Tab:eff_g}.

\subsection{Polarizability coefficient}
The polarizability coefficient ($W_{2}$) is a scaling factor which quantifies the fraction of linear polarization produced by resonant scattering of the incoming radiation. Analytical expressions for $W_{2}$ for three allowed transitions corresponding to $\Delta J = J_2 - J_1 = 0, \pm 1$ (with $J_{2} = J_{1} = 0$ being forbidden) are given by \citep{Stenflo1994ASSL..189.....S}

Case I: When $J_{2} = J_{1} - 1$,
\begin{equation}
\label{eq:C1}
    W_{2} = \frac{(J_{1} - 1)(2J_{1} - 3)}{10J_{1}(2J_{1} + 1)}
\end{equation}

Case II: When $J_{2} = J_{1}$,
\begin{equation}
\label{eq:C2}
    W_{2} = \frac{(2J_{1} - 1)(2J_{1} + 3)}{10J_{1}(J_{1} + 1)}
\end{equation}

Case III: When $J_{2} = J_{1} + 1$,
\begin{equation}
\label{eq:C3}
    W_{2} = \frac{(J_{1} + 2)(2J_{1} + 5)}{10(J_{1} + 1)(2J_{1} + 1)}
\end{equation}

The calculated values of $W_{2}$ for the corresponding UV spectral lines are listed in the tenth column of Table~\ref{Tab:eff_g}.

\subsection{Line irradiance}
For this work, only the coronal lines are focused on. The EIS off-limb spectra consists of spectral lines whose intensities over active region and quiet sun have been collected from \cite{Zanna2012A&A...537A..38D}. The off-limb line intensities from 470 to 1609~\AA ~spanning over three regions (quiet sun (QS), active region (AR) and coronal hole (CH)) have been assembled from \cite{curdt2004sumer}. All the line irradiances from EIS and SUMER are available in the units of photons cm$^{-2}$ s$^{-1}$ arcsec$^{-2}$. But for the EUNIS on-disk observations of spectral lines from 300 to 370 \AA, the given line irradiances are converted from erg cm$^{-2}$ s$^{-1}$ sr$^{-1}$ to photons cm$^{-2}$ s$^{-1}$ arcsec$^{-2}$ (formula \ref{eq:int_co}) in order to simplify the comparison with other spectral lines (from EIS and SUMER observations) used in the present analysis. The spectral irradiances for different solar regions, where available, are listed in Table~\ref{Tab:eff_g}  in the units of photons cm$^{-2}$s$^{-1}$arcsec$^{-2}$.

 \begin{equation}
  \label{eq:int_co}
    1 ~erg ~cm^{-2} s^{-1} sr^{-1} = 11.8324 \times 10^{-4} \times \lambda ~photons ~cm^{-2} s^{-1} arcsec^{-2}
\end{equation}
where, $\lambda$ is the wavelength (in \AA) of the corresponding spectral line.

\subsection{Intensity derivative}
In the context of inferring vector magnetic field in the corona, the expected amplitude of Stokes $V$ profiles of a few selected UV lines have been estimated in this work. In addition to the effective Land\'e factor and the wavelength, the Stokes $V$ signal also depends on the intensity derivative $\frac{dI}{d\lambda}$ (cf. equation \ref{eq:Vsignal}). The spectroscopic observations from SUMER (for details, refer to \citealp{curdt2001spectral, curdt2004sumer}) are used for selecting the spectral lines.
Relevant data have been downloaded from \url{https://sdac.virtualsolar.org}. 

\begin{figure}[htbp]
   \centerline{\hspace*{0.01\textwidth}
               \includegraphics[width=0.35\textwidth,clip=]{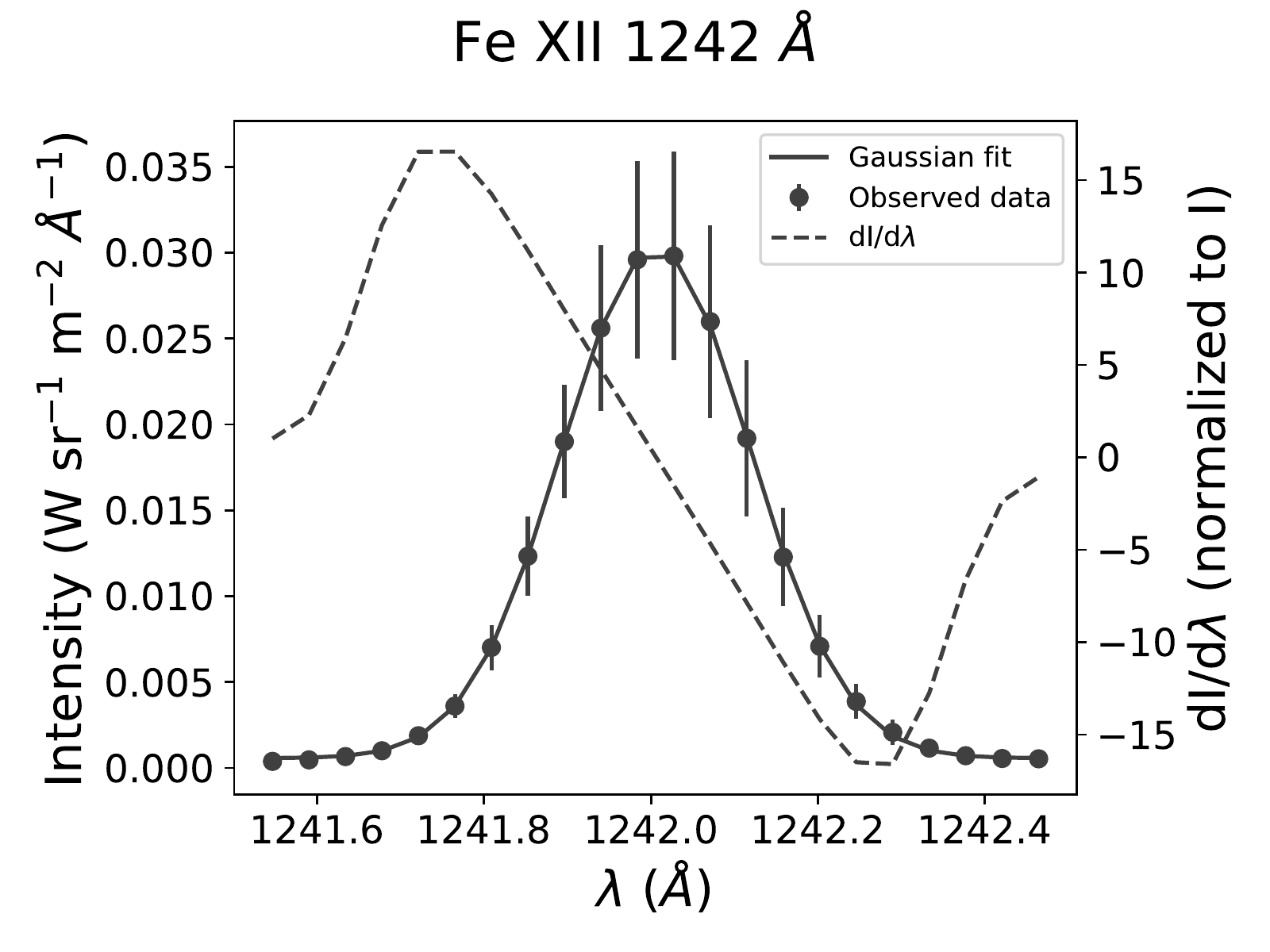}
               \hspace*{-0.018\textwidth}
               \includegraphics[width=0.35\textwidth,clip=]{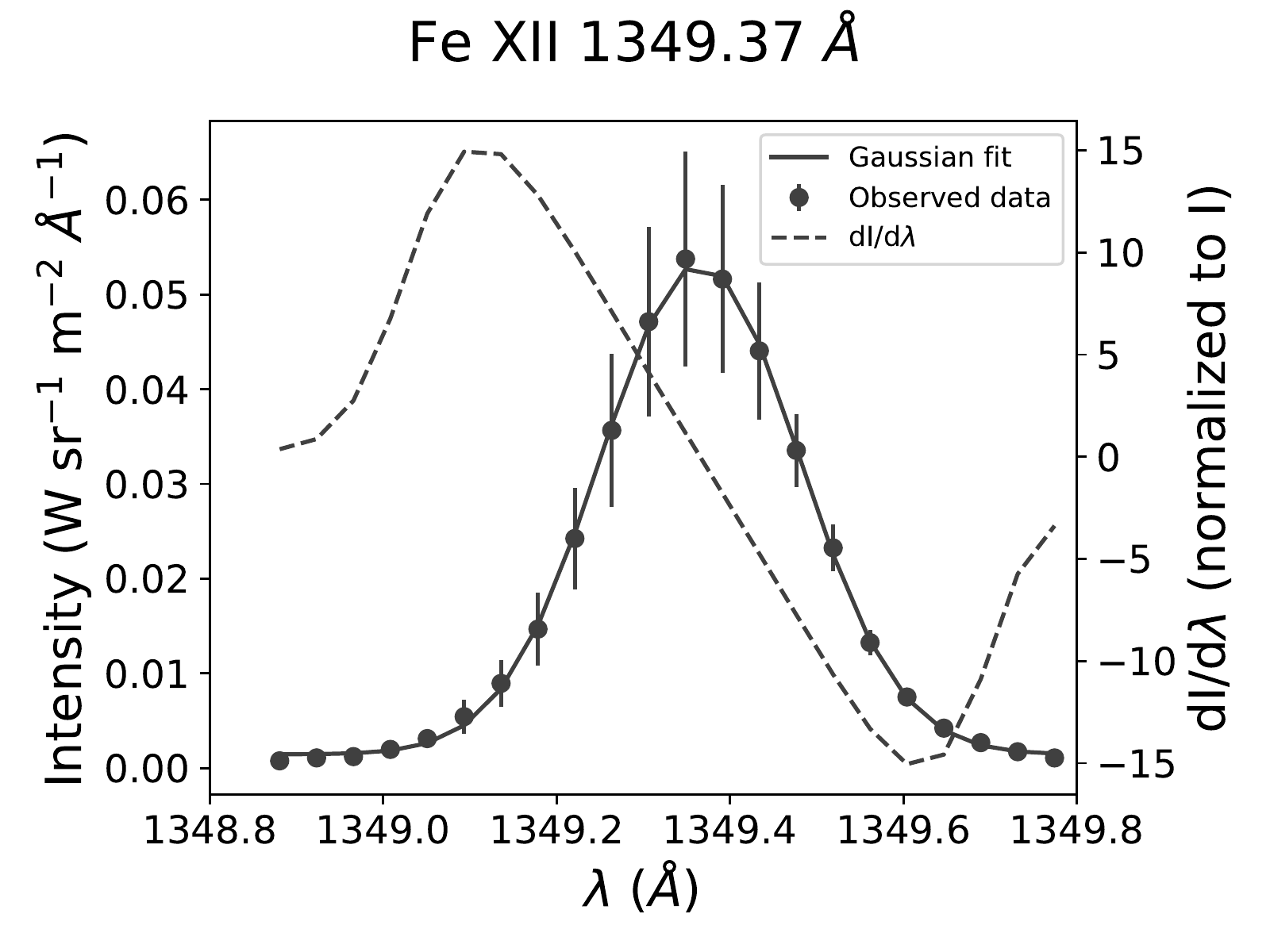}
               \hspace*{-0.018\textwidth}
               \includegraphics[width=0.35\textwidth,clip=]{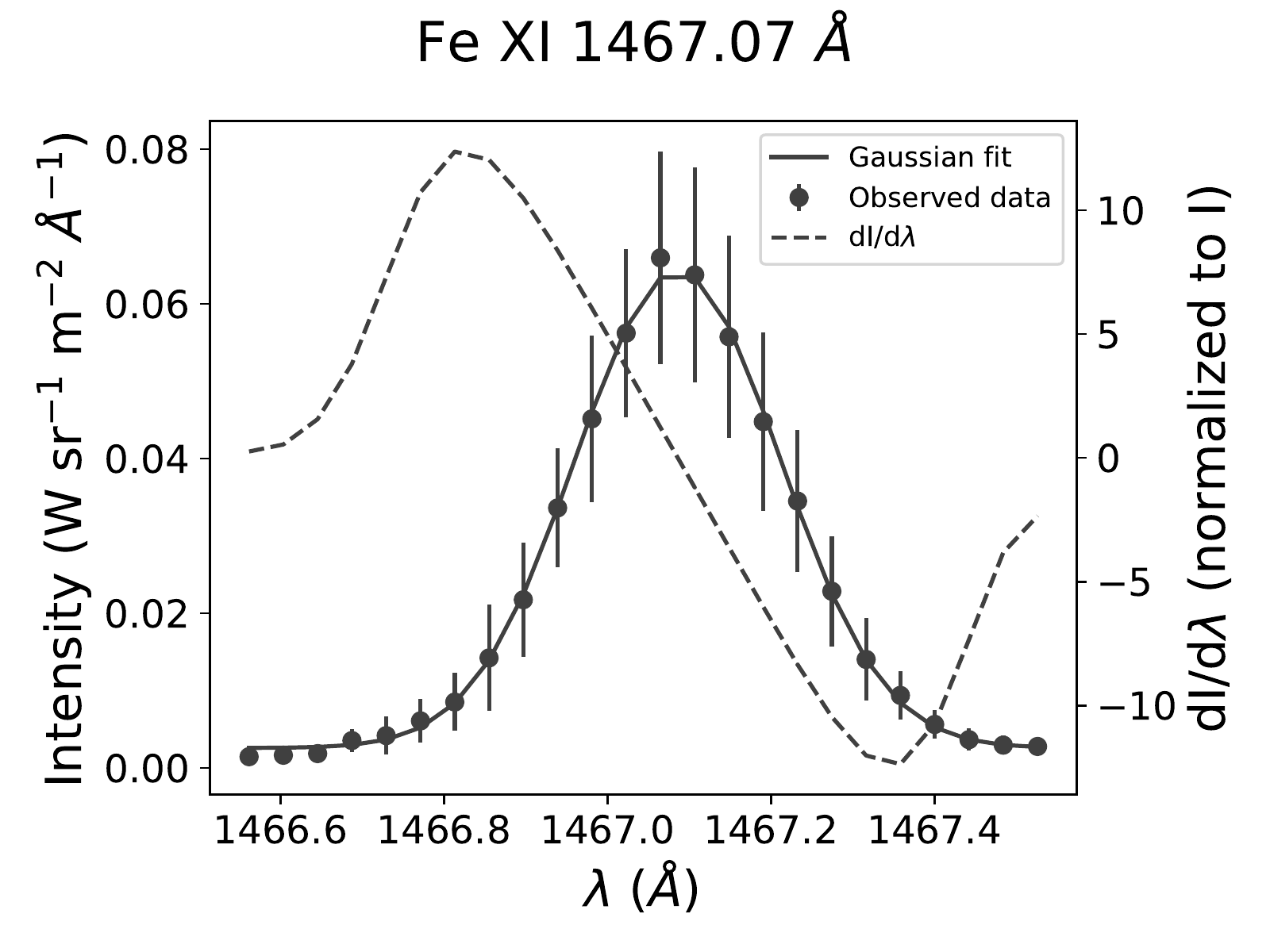}
              }
\caption{Sample intensity profiles of Fe~{\sc xii} at 1242~\AA (the left panel) and  1349~\AA (the middle panel), and Fe~{\sc xi} at 1467~\AA (the right panel) are shown along with their corresponding derivative plots. The solid curves are the Gaussian fit to the observed data points (the filled circles). The error bars correspond to the dispersion in intensity values across the spatial pixels over which the spectral profiles are averaged. The dotted curves are the derivative of the fitted Gaussian curves normalized to the local intensities.}
   \label{Fig:Int_prof}
\end{figure}

Before estimating the intensity derivative, the spectral profiles corresponding to a given spectral line are fitted with Gaussian function. The derivatives of intensity with respect to wavelength are calculated and normalized to the local intensity of each Gaussian fitted profile. Then the derivative values (absolute and normalized) at half-maxima of both the blue and the red wings are averaged using which the Stokes~$V/I$ amplitude is estimated for each Gaussian fitted profile along the spatial axis via equation \ref{eq:Vsignal}. Finally, the mean Stokes~$V/I$ amplitude is calculated for the given spectral line. The standard deviation of the Stokes~$V/I$ amplitude is also estimated relative to its mean value over the fitted spectral profiles. In the above analysis, the Stokes $V/I$ amplitude increases by 1.3 (for detector A) and 1.1 (for detector B) when the Stokes I profiles are corrected for the instrumental broadening. This implies that there is no significant change in the Stokes $V/I$ amplitudes due to instrumental broadening effect and therefore while calculating Stokes $V/I$, it can safely be unaccounted for. 

Figure~\ref{Fig:Int_prof} shows the sample mean intensity profiles of a few selected forbidden lines in FUV due to Fe~{\sc xii} at 1242~\AA~(the left panel), 1349~\AA~(the middle panel), and Fe~{\sc xi} at 1467~\AA~(the right panel) along with their corresponding mean intensity derivatives (the dotted curves). The error bars correspond to the dispersion in intensity values across the spatial pixels over which the spectral profiles are averaged. For a LOS field of 10~Gauss, the expected Stokes~$V/I$ amplitudes for Fe~{\sc xii} lines at 1242~\AA ~and  1349~\AA \ are $(1.22 \pm 0.096)\times10^{-4}$  and $(1.95 \pm 0.069)\times10^{-4}$, respectively, and for Fe~{\sc xi} line at 1467~\AA, is $(1.34 \pm 0.11)\times10^{-4}$. Similar calculations have been done for spectral lines below 1200 \AA ~and we found that the Stokes~$V/I$ signal is in the order of 10$^{-5}$ or less.

\section{Summary and Conclusion}
\label{Sec:Res_Disc}

\begin{figure}[htbp]
       \centering
      \includegraphics[width=1\textwidth,clip=]{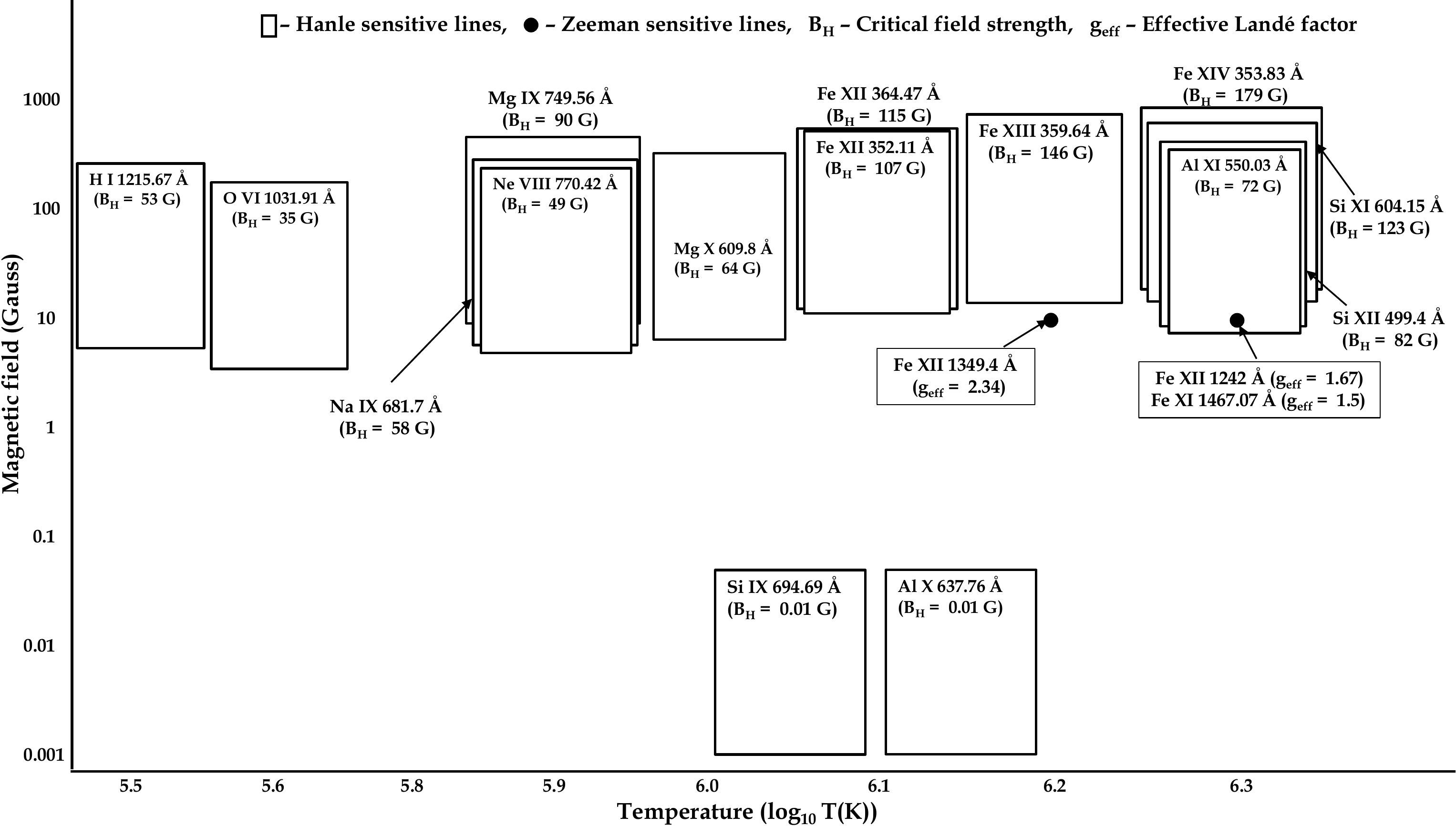}
    \caption{Graphical representation of magnetic sensitivity of the spectral lines in the domain of Hanle effect (cf. Eq. \ref{eq:HE_cond3}). The domains are shown in rectangular boxes, each of which covers an approximate temperature range along the X-axis and, magnetic field strength from 0.1B$_{H}$ to 5B$_{H}$ along the Y-axis (Not to be scaled). The middle of each rectangular box along the X-axis respresents the peak line formation temperature. FUV spectral lines, having best estimated Stokes $V$ signal, are shown using solid circles at a coronal field of 10 G.}
    \label{fig:Hanle_Zee_sens} 
\end{figure}

The search for spectral lines in FUV and EUV to probe vector magnetic field has been carried out in this paper. The outcome of this search is summarized in a graphical representation shown in Figure~\ref{fig:Hanle_Zee_sens}. In this figure each line is represented by a rectangular box with its length along Y-axis indicating the magnetic field sensitivity range due to Hanle effect as dictated by Eq.~\ref{eq:HE_cond3}. The width of the box along X-axis is not a true representation of temperature sensitivity range but only to indicate the peak formation temperature given in logarithmic scale. Actual temperature sensitivity extends much beyond that is indicated by the width of the box. Each box is marked by the wavelength of the spectral line along with the corresponding name of the ion and Hanle critical magnetic field ($B_H$). The closed solid circles in this figure represent the FUV lines shown in Figure \ref{Fig:Int_prof}. Their locations in Figure~\ref{fig:Hanle_Zee_sens} indicate the peak formation temperature on the X-axis and the assumed field strength of 10 Gauss, at which their Stokes~$V/I$ signal is expected to be in the order of $10^{-4}$, on the Y-axis. For estimation of Stokes~$V/I$ signal, Eq.~\ref{eq:Vsignal} is used along with the measured intensity derivative ($\frac{dI}{d\lambda}$). While selecting the spectral lines presented in Figure~\ref{fig:Hanle_Zee_sens}, the polarizability coefficient~($W_2$), line irradiance and $B_H$ are considered. Only lines with $B_H$ in the range 0.01 - 200 G, $W_2>0$ and their intensity $>$ 1~photon~cm$^{-2}$s$^{-1}$arcsec$^{-2}$ are shown in Figure~\ref{fig:Hanle_Zee_sens}. The range of field strength chosen is directed by the coronal magnetic field measurements reported in the literature (see for e.g., Fig 5. of \citealp{Peter2012ExA....33..271P} and Figure 4. of \citealp{Sasikumar2021arXiv211014179S}).

Regarding the intensity criteria, it is apparent that the spectral lines with maximum irradiance should be chosen as, spectropolarimetric observations are photon starved. However, it is difficult to find spectral lines with high irradiance which are sensitive to Hanle and Zeeman effects and, cover a suitable temperature range as well. The O~{\sc vi} line at 1031.91~\AA ~is a good Hanle sensitive line both in terms of number of photons and $B_H$. However, in order to derive vector magnetic field there is no other spectral line with the same peak formation temperature. Given the fact that the line formation temperature is not a delta function but has a range, this line can be used with other lines having adjacent formation temperature. For instance, this line can be used together with Ne~{\sc viii} at 770.42~\AA ~to derive vector magnetic field information. Though the number of photons in these two lines are relatively close, they are separated in their wavelengths by $\approx 260$~\AA ~and formation temperature differs by $\approx 0.3$ on logarithmic scale. Similarly, Ly-$\alpha$ line at 1215.67 \AA~is sensitive to Hanle effect (B$_H$= 53 G) having extremely high line irradiance. However, there is no spectral line with similar line irradiance and formation temperature to be used in association with Ly-$\alpha$.

The O~{\sc vi} at 1037.61~\AA ~and Ne~{\sc viii} at 780.39~\AA~lines with $W_2=0$ can be used for zero polarization reference which may help in correcting for systematic artifacts. The Na~{\sc ix} at 681.72~\AA ~can also be used together with Ne~{\sc viii} at 770.42~\AA ~but the number of photons is significantly less (close to a factor of 4). The combination of Ne~{\sc viii} at 770.42~\AA ~($\mathrm{log}_{10}(T)=5.9$, $B_H=49$~G) and Mg~{\sc x} at 609.79~\AA ~($\mathrm{log}_{10}(T)=6.0$, $B_H=64$~G) is good to probe vector magnetic field from the regions of plasma with temperature in the range $\mathrm{log}_{10}(T)=5.9-6.0$. Both lines have good line irradiance. However, the wavelength separation is about 160~\AA~between these two lines. Nevertheless, with this line combination also there are two spectral lines viz., Ne~{\sc viii} at 780.39~\AA ~and Mg~{\sc x} at 624.94~\AA, which can be used as zero polarization reference. Interestingly, Si~{\sc ix} at 694.69 \AA~has similar line irradiance as Na~{\sc ix} at 681.72 \AA~and exhibit different sensitivity to Hanle effect, with ${g_{u}} \omega_{B}\tau =$ 1 for 0.01 G and 58 G, respectively. Therefore, Si~{\sc ix} 694.69 \AA~line will be principally sensitive to the field direction, while Na~{\sc ix} 681.72 \AA~will be suited for determining the magnetic field strength.

There are several spectral lines in the wavelength range of 350 - 370 \AA ~with B$_H=100-180$ G which can be used to probe vector field from the regions with plasma temperature of $\mathrm{log}_{10}(T)=$ 6.1 to 6.3. The greatest advantage here is that these lines are clustered within a wavelength band of about 20~\AA~which is beneficial in terms of instrument design and development. However, these lines have moderate number of photons compared to the lower temperature lines mentioned in the preceding paragraphs. At $\mathrm{log}_{10}(T)=6.3$ there are two spectral lines which are best suited for vector magnetic field measurements viz., Al~{\sc xi} at 550~\AA ~and Si~{\sc xii} at 499.4~\AA ~both in terms of their wavelength proximity and $B_H$ values which are 72~G and 82~G, respectively.

Some of the Hanle saturated lines in FUV are also explored in the context of providing additional constraints on the LOS component of the magnetic field vector. However, their wavelength separation is larger, with respect to Hanle sensitive lines of identical formation temperatures, making them less attractive for vector magnetic field measurements (cf. Figure~\ref{fig:Hanle_Zee_sens}). Hence the spectral lines with different Hanle sensitivity are best suited for probing vector magnetic field in the solar corona, at least in the temperature range of $\mathrm{log}_{10}(T)=5.5-6.3$. 
The Hanle sensitivity of the spectral lines given in Figure~\ref{fig:Hanle_Zee_sens} is limited to $\ge 4$~Gauss, with the exception of Si {\sc ix} and Al {\sc x} lines having Hanle sensitivity from 0.001 to 0.05 Gauss which may be useful in probing very weak magnetic field in the milliGauss range. This implies that the coronal height up to which most of the magnetic field measurements can be carried out is limited to $<2$~R$_{\odot}$. At this range of height, the impact of electron collisions becomes significant due to larger electron density and one of its main effect is depolarization. In order to estimate the depolarization factor due to collisions, it is important to calculate and compare between the collisional and the radiative rates through detailed modeling of each individual line shown in Figure \ref{fig:Hanle_Zee_sens}. Therefore, depolarizing collisions along with other symmetry breaking processes such as non-radial solar wind, ion temperature anisotropy and presence of active regions \citep{Fineschi1993SPIE.1742..423F, Zhao2019ApJ...883...55Z, Zhao2021ApJ...912..141Z} must be taken into account while interpreting the spectropolarimetric information in actual observations. 


\begin{sidewaystable}[htbp]        
\centering

\caption{Table showing ionization state of the atomic species, corresponding wavelength and level configuration, temperature of line formation (log$_{10}$(T)), type of transition (TT: E1 or M1), effective Land\'e factor ($\overline{g}$) , critical field strength (B$_{H}$), polarizability coefficient (W$_{2}$) and line irradiances in three different solar regions. E1 and M1 refer to electric dipole and magnetic dipole transitions, respectively. $^{2S+1}$L$_J$ is the notation used in each atomic level configuration, where L, S and J are total orbital, spin and angular momentum quantum numbers respectively.}
    \label{Tab:eff_g}
\begin{adjustbox}{width= 1\textwidth}

\begin{threeparttable}
\begin{tabular}{llllclllccccc}     
\hline                   
  
\textbf{Ion} & \textbf{$\lambda$(\AA)} & \textbf{Transition (i-k)} & \textbf{log$_{10}$(T)} & \textbf{TT} & \bf{$A(s^{-1})$} &  $\bf{g_{i}-g_{k}}$ & $\bf{\overline{g}}$ & \textbf{B$_{H}$} & \textbf{W$_{2}$} & \multicolumn{3}{c}{\textbf{L \tiny (ph cm$^{-2}$s$^{-1}$arcsec$^{-2}$)}} \\
  &      &     &   &     &       &   &      &  \textbf{(Gauss)} &   & \textbf{QS} & \textbf{AR} & \textbf{CH} \\
  \hline
Fe IX & 171.073 & 3s$^{2}$ 3p$^{6}$ $^{1}$S$_{0}$  & 6.11 & E1 & $2.28\times10^{11}$ & 1 - 1 & 1  & 25817 & 1  & 140.7 & 265 & - \\
      &         & - 3s$^{2}$ 3p$^{5}$ 3d $^{1}$P$_{1}$ &         &           &    &   &  & \\
    \hline
Fe X & 174.531 & 3s$^{2}$ 3p$^{5}$ $^{2}$P$_{3/2}$ & 6 & E1 & $1.86\times10^{11}$ & 1.33-1.2 & 1.1 & 17551  & 0.28 & 119.7 & 172.7 & - \\
      &         & - 3s$^{2}$ 3p$^{4}$ 3d $^{2}$D$_{5/2}$ &         &           &       &   &\\
    \hline
Fe X & 184.537 & 3s$^{2}$ 3p$^{5}$ $^{2}$P$_{3/2}$ &  6 & E1  & $1.63\times10^{11}$  & 1.33 - 2 & 1.17 &  9245 & 0 & 33.6 & 47.9 & - \\
      &         & - 3s$^{2}$ 3p$^{4}$ 3d $^{2}$S$_{1/2}$ &         &           &       &   & \\
    \hline
Fe XII & 195.119 & 3s$^{2}$ 3p$^{3}$ $^{4}$S$_{3/2}$ & 6.1 & E1 & $8.83\times10^{10}$  & 2 - 1.6 & 1.3 &  6249 & 0.28 & 136.2 & 312 & - \\
      &         & - 3s$^{2}$ 3p$^{2}$ 3d $^{4}$P$_{5/2}$ &         &           &       &   & \\
    \hline
Fe XIII & 202.044 & 3s$^{2}$ 3p$^{2}$ $^{3}$P$_{0}$ & 6.2 & E1 & $5.33\times10^{10}$   & 1.5 - 1.5 & 1.5 &  4024  & 1 & 103.3 & 349.5 & - \\
      &         & - 3s$^{2}$ 3p 3d $^{3}$P$_{1}$ &         &           &       &   & \\
    \hline
Si X & 258.374 & 2s$^{2}$ 2p $^{2}$P$_{3/2}$  & 6.1 & E1  & $1.72\times10^{10}$  & 1.33-1.33 & 1.33 & 1461 & 0.32 & 44.1 & 66.3 & - \\
      &         & - 2s 2p$^{2}$ $^{2}$P$_{3/2}$ &         &           &       & \\
    \hline
Si X & 261.056 & 2s$^{2}$ 2p $^{2}$P$_{3/2}$  & 6.1 & E1  & $1.68\times10^{10}$  & 1.33-0.67 & 1.5 & 2853 & 0  & 26.3 & 37.8 & -  \\
      &         & - 2s 2p$^{2}$ $^{2}$P$_{1/2}$ &         &           &       &   & \\
    \hline
Fe XIV & 264.788 & 3s$^{2}$ 3p $^{2}$P$_{3/2}$  & 6.3 & E1  & $4.26\times10^{10}$  & 1.33-1.33 & 1.33 &  3618 & 0.32 & 17.1 & 92.1 & -\\
      &         & - 3s 3p$^{2}$ $^{2}$P$_{3/2}$ &         &           &       &    & \\
    \hline
Fe XV  & 284.163 & 3s$^{2}$ $^{1}$S$_{0}$  & 6.3 & E1  & $2.11\times10^{10}$  & 1 - 1 & 1 &  2389 & 1 & 24.3 & 576.6 & - \\
      &         & - 3s 3p $^{1}$P$_{1}$   &         &           &       &     & \\
    \hline
Si XI & 303.325  & 1s$^{2}$ 2s$^{2}$ $^{1}$S$_{0}$  & 6.2 & E1  & $6.38\times10^{9}$  & 1 - 1 & 1 & 722 & 1 & - & 1049 & - \\
      &         & -  1s$^{2}$ 2s 2p $^{1}$P$_{1}$ &         &           &       &    &  \\
    \hline
Fe XI & 308.544  & 3s$^{2}$ 3p$^{4}$ $^{1}$D$_{2}$  & 6.1 & E1  & $8.84\times10^{9}$  & 1 - 1 & 1 & 1001  & 0.01 & - & 13 & - \\
      &         & - 3s 3p$^{5}$ $^{1}$P$_{1}$ &         &           &       &    &  \\
    \hline
Fe XIII & 312.174  & 3s$^{2}$ 3p$^{2}$ $^{3}$P$_{1}$ & 6.2 & E1  & $4.07\times10^{9}$  & 1.5 - 1.5 & 1.5 & 307 & 0.25 & - & 12 & -  \\
      &         & - 3s 3p$^{3}$ $^{3}$P$_{1}$  &         &           &       &     &  \\
    \hline
\end{tabular}
\end{threeparttable}

\end{adjustbox}
\end{sidewaystable}

\begin{sidewaystable}[htbp]
\begin{turn}{180}
\begin{adjustbox}{width= 1\textwidth}

\centering
\begin{threeparttable}
\begin{tabular}{llllclllccccc}     
\hline                   
  
\textbf{Ion} & \textbf{$\lambda$(\AA)} & \textbf{Transition (i-k)} &\textbf{log$_{10}$(T)} & \textbf{TT} & \bf{$A(s^{-1})$} &  $\bf{g_{i}-g_{k}}$ & $\bf{\overline{g}}$ & \textbf{B$_{H}$} & \textbf{W$_{2}$} & \multicolumn{3}{c}{\textbf{L \tiny (ph cm$^{-2}$s$^{-1}$arcsec$^{-2}$)}} \\
  &      &      &   &     &      &   &      &  \textbf{(Gauss)}  &  & \textbf{QS} & \textbf{AR} & \textbf{CH} \\
  \hline
Fe XIII & 318.13  & 3s$^{2}$ 3p$^{2}$ $^{1}$D$_{2}$ & 6.2 & E1  &  $5.61\times10^{9}$  & 1 - 1 & 1 & 635 & 0.35  & -  & 10.8 & - \\
      &         & - 3s 3p$^{3}$ $^{1}$D$_{2}$  &         &           &       &    & \\
    \hline
Fe XIII & 320.8  & 3s$^{2}$ 3p$^{2}$ $^{3}$P$_{2}$  & 6.2 & E1 & $3.89\times10^{9}$  & 1.5 - 1.5 & 1.5 & 294 & 0.35  & - & 40.4  & - \\
      &         & - 3s 3p$^{3}$ $^{3}$P$_{2}$  &         &           &       &    & \\
    \hline
Fe XV & 327.033  & 3s 3p $^{3}$P$_{2}$  & 6.3 & E1  & $4.67\times10^{9}$  & 1.5 - 1 & 1.25 & 529 & 0.35  & - & 6.4 & - \\
      &         & - 3p$^{2}$ $^{1}$D$_{2}$ &         &           &       &       & \\
    \hline
Cr XIII & 328.268 & 2p$^{6}$ 3s$^{2}$ $^{1}$S$_{0}$  & 6.2 & E1 & $1.73\times10^{10}$  & 1 - 1 & 1 & 1959 & 1  & - & 25 & - \\
      &         & - 3s 3p $^{1}$P$_{1}$  &         &           &       &    & \\
    \hline
Al X & 332.79 & 1s$^{2}$ 2s$^{2}$ $^{1}$S$_{0}$  & 6.1 & E1  & $5.74\times10^{9}$  & 1 - 1 & 1 & 650 & 1  & - & 35.2  & - \\
      &         & - 1s$^{2}$ 2s 2p $^{1}$P$_{1}$ &         &           &       &    & \\
    \hline
Fe XIV & 334.178 & 3s$^{2}$ 3p $^{2}$P$_{1/2}$  & 6.3 & E1  & $2.49\times10^{9}$  & 0.67 - 0.8 & 0.83 & 352 & 0.5  & - & 74 & -  \\
        &         & - 3s 3p$^{2}$ $^{2}$D$_{3/2}$  &         &           &       &   & \\
    \hline
Fe XVI & 335.409 & 2p$^{6}$ 3s $^{2}$S$_{1/2}$ & 6.4 & E1  & $7.87\times10^{9}$  & 2 - 1.33 &  1.17 & 669 & 0.5  & - & 159 & - \\
        &         & - 2p$^{6}$ 3p $^{2}$P$_{3/2}$  &         &           &       &     & \\
    \hline
Fe XII & 338.263 & 3s$^{2}$ 3p$^{3}$ $^{2}$D$_{5/2}$  & 6.1 & E1  & $3.37\times10^{9}$  & 1.2 - 1.2 & 1.2 & 318 & 0.37  & - & 23.8 & - \\
        &         & - 3s 3p$^{4}$ $^{2}$D$_{5/2}$  &         &           &       &   & \\
    \hline
Fe XI & 341.113 & 3s$^{2}$ 3p$^{4}$ $^{3}$P$_{2}$  & 6.1 & E1  & $3.28\times10^{9}$  & 1.5 - 1.5 & 1.5 & 248 & 0.01 & - & 25.5 & - \\
        &         & - 3s 3p$^{5}$ $^{3}$P$_{1}$  &         &           &       &   & \\
    \hline
Si IX & 341.949 & 2s$^{2}$ 2p$^{2}$ $^{3}$P$_{0}$  & 6.1 & E1  & $2.4\times10^{9}$  & 1.5 - 0.5 & 1 & 544 & 1  & - & 18.2 & -\\
          &         & - 2s 2p$^{3}$ $^{3}$D$_{1}$  &         &           &       &     &\\
    \hline
Fe XII & 346.852 & 3s$^{2}$ 3p$^{3}$ $^{4}$S$_{3/2}$  & 6.1 & E1  & $1.86\times10^{9}$  & 2 - 2.67 & 1.83 & - & 0  & - & 30.2 & -  \\
            &         & - 3s 3p$^{4}$ $^{4}$P$_{1/2}$  &         &           &       &       & \\
    \hline
Fe XIII & 348.183 & 3s$^{2}$ 3p$^{2}$ $^{3}$P$_{0}$  & 6.2 & E1  & $1.61\times10^{9}$  & 1.5 - 0.5 & 1 & 365 & 1 & - & 29.3 & - \\
            &         & - 3s 3p$^{3}$ $^{3}$D$_{1}$  &         &           &       &      &  \\
    \hline
Fe XII & 352.106 & 3s$^{2}$ 3p$^{3}$ $^{4}$S$_{3/2}$  & 6.1 & E1  & $1.64\times10^{9}$  & 2 - 1.73 & 1.87 & 107 & 0.32 & - & 64.8 & - \\
            &         & - 3s 3p$^{4}$ $^{4}$P$_{3/2}$  &         &           &       &       & \\
    \hline
\end{tabular}
\end{threeparttable}

\end{adjustbox}
\end{turn}
\end{sidewaystable}

\begin{sidewaystable}[htbp]
\begin{adjustbox}{width= 1\textwidth}
\centering

\begin{threeparttable}
\begin{tabular}{llllclllccccc}     
\hline                   
  
\textbf{Ion} & \textbf{$\lambda$(\AA)} & \textbf{Transition (i-k)} &\textbf{log$_{10}$(T)} & \textbf{TT} & \bf{$A(s^{-1})$} &  $\bf{g_{i}-g_{k}}$ & $\bf{\overline{g}}$ & \textbf{B$_{H}$} & \textbf{W$_{2}$} & \multicolumn{3}{c}{\textbf{L \tiny (ph cm$^{-2}$s$^{-1}$arcsec$^{-2}$)}} \\
  &      &      &   &      &      &   &      &  \textbf{(Gauss)}  &  & \textbf{QS} & \textbf{AR} & \textbf{CH} \\
  \hline
Fe XI & 352.67 & 3s$^{2}$ 3p$^{4}$ $^{3}$P$_{2}$  & 6.1 & E1  & $3.11\times10^{9}$ & 1.5 - 1.5 & 1.5 & 235 & 0.35  & - & 64.3 & - \\
            &         & - 3s 3p$^{5}$ $^{3}$P$_{2}$  &         &           &       &      &  \\
    \hline
Fe XIV & 353.829 & 3s$^{2}$ 3p $^{2}$P$_{3/2}$  & 6.3 & E1  & $1.9\times10^{9}$  & 1.33 - 1.2 & 1.1 & 179 & 0.28 & - & 38.6 & - \\
            &         & - 3s 3p$^{2}$ $^{2}$D$_{5/2}$ &         &           &       &      &  \\
    \hline
Fe XI & 356.519 & 3s$^{2}$ 3p$^{4}$ $^{3}$P$_{1}$  & 6.1 & E1  & $3.28\times10^{9}$ & 1.5 - 1.5 & 1.5 & 248 & 0.25 & - & 10.4 & -  \\
            &         & - 3s 3p$^{5}$ $^{3}$P$_{1}$  &         &           &       &      &  \\
    \hline
Fe XIII & 359.644 & 3s$^{2}$ 3p$^{2}$ $^{3}$P$_{1}$  & 6.2 & E1 & $1.5\times10^{9}$  & 1.5 - 1.17 & 1 & 146 & 0.35 & - & 54.1 & - \\
            &         & - 3s 3p$^{3}$ $^{3}$D$_{2}$  &         &           &       &       & \\
    \hline
Fe XIII & 359.839 & 3s$^{2}$ 3p$^{2}$ $^{3}$P$_{1}$  & 6.2 & E1  & $1.61\times10^{9}$ & 1.5 - 0.5 & 1 & 365 & 0.25 & - & 5.3 &  - \\
            &         & - 3s 3p$^{3}$ $^{3}$D$_{1}$  &         &           &       &      &  \\
    \hline
Fe XVI & 360.758 & 2p$^{6}$ 3s $^{2}$S$_{1/2}$  & 6.4 & E1  & $6.34\times10^{9}$ & 2 - 0.67 & 1.33 & - & 0 & - & 76.9 & - \\
            &         & - 2p$^{6}$ 3p $^{2}$P$_{1/2}$  &         &           &       &       & \\
    \hline
Fe XII & 364.467 & 3s$^{2}$ 3p$^{3}$ $^{4}$S$_{3/2}$  & 6.1 & E1  & $1.62\times10^{9}$ & 2 - 1.6 & 1.33 & 115 & 0.28 & - & 85.4 & - \\
            &         & - 3s 3p$^{4}$ $^{4}$P$_{5/2}$  &         &           &       &       & \\
    \hline
Mg IX & 368.071 & 1s$^{2}$ 2s$^{2}$ $^{1}$S$_{0}$  & 6.0 & E1  & $5.12\times10^{9}$  & 1 - 1 & 1 & 580 & 1  & - & 486 & - \\
            &         & - 1s$^{2}$ 2s 2p  $^{1}$P$_{1}$  &         &           &       &       & \\
    \hline
Fe XI & 369.163 & 3s$^{2}$ 3p$^{4}$ $^{3}$P$_{1}$  & 6.1 & E1  & $3.11\times10^{9}$ & 1.5 - 1.5 & 1.5 & 235 & 0.35 & - & 20.5 & - \\
            &         & - 3s 3p$^{5}$ $^{3}$P$_{2}$  &         &           &       &       & \\
    \hline
Si XII & 499.4 & 1s$^{2}$ 2s $^{2}$S$_{1/2}$  & 6.3 & E1  & $9.61\times10^{8}$  & 2 - 1.33 & 1.17 & 82 & 0.5 & 23.3 & 166.4 & - \\
            &         & - 1s$^{2}$ 2p $^{2}$P$_{3/2}$  &         &           &       &       & \\
   \hline
Si XII & 520.67 & 1s$^{2}$ 2s $^{2}$S$_{1/2}$  & 6.26 & E1  & $8.48\times10^{8}$  & 2 - 0.67 & 1.33 & - & 0 & 13.3 & 100.9 & -  \\
            &         & - 1s$^{2}$ 2p $^{2}$P$_{1/2}$ &         &           &       &      &  \\
   \hline
Al XI & 550.031 & 1s$^{2}$ 2s $^{2}$S$_{1/2}$  & 6.26 & E1  & $8.52\times10^{8}$  & 2 - 1.33 & 1.17 & 72 & 0.5 & 7.9 & 48.5 & - \\
            &         & - 1s$^{2}$ 2p $^{2}$P$_{3/2}$  &         &           &       &      &  \\
   \hline
Ca X  & 557.76 & 2p$^{6}$ 3s $^{2}$S$_{1/2}$  & 5.85 – 6 & E1   & $3.77\times10^{9}$  & 2 - 1.33 & 1.17 & 320 & 0.5  & 9.0 & 19.7 & 5.2 \\
            &         & - 2p$^{6}$ 3p $^{2}$P$_{3/2}$  &         &           &       &      &  \\
    \hline
\end{tabular}
\end{threeparttable}

\end{adjustbox}
\end{sidewaystable}

\begin{sidewaystable}[htbp]
\begin{turn}{180}
\begin{adjustbox}{width= 1\textwidth}
\centering

\begin{threeparttable}
\begin{tabular}{llllclllccccc}     
\hline                   
  
\textbf{Ion} & \textbf{$\lambda$(\AA)} & \textbf{Transition (i-k)} &\textbf{log$_{10}$(T)} & \textbf{TT} & \bf{$A(s^{-1})$} &  $\bf{g_{i}-g_{k}}$ & $\bf{\overline{g}}$ & \textbf{B$_{H}$} & \textbf{W$_{2}$} & \multicolumn{3}{c}{\textbf{L \tiny (ph cm$^{-2}$s$^{-1}$arcsec$^{-2}$)}} \\
  &      &      &   &      &      &   &      &  \textbf{(Gauss)}  &   & \textbf{QS} & \textbf{AR} & \textbf{CH} \\
  \hline
Al XI & 568.12 & 1s$^{2}$ 2s $^{2}$S$_{1/2}$  & 6.26 & E1   & $7.73\times10^{8}$ & 2 - 0.67 & 1.33 & - & 0 & 5.3 & 23.5 & - \\
            &         & - 1s$^{2}$ 2p $^{2}$P$_{1/2}$  &         &           &       &      &  \\
    \hline
Ca X  & 574.01 & 2p$^{6}$ 3s $^{2}$S$_{1/2}$  & 5.85 – 6 & E1  & $3.47\times10^{9}$ & 2 - 0.67 & 1.33 & - & 0 & 6.0 & 12.2 & 3.2 \\
            &         & - 2p$^{6}$ 3p $^{2}$P$_{1/2}$  &         &           &       &      &  \\
    \hline
Si XI  & 604.15 & 2s 2p $^{1}$P$_{1}$  & 6.26 & E1  & $1.09\times10^{9}$  & 1 - 1 & 1 & 123 & 0.35  & 2.5 & 4.1 & - \\
            &         & - 2p$^{2}$ $^{1}$D$_{2}$  &         &           &       &      &  \\
    \hline
Mg X & 609.793 & 1s$^{2}$ 2s $^{2}$S$_{1/2}$  & 6.04 & E1  & $7.53\times10^{8}$  & 2 - 1.33 & 1.17 & 64 & 0.5 &  213 & 341 & 26 \\
            &         & - 1s$^{2}$ 2p $^{2}$P$_{3/2}$ &         &           &       &      &  \\
    \hline
Mg X & 624.94 & 1s$^{2}$ 2s $^{2}$S$_{1/2}$  & 6 & E1  & $7\times10^{8}$ & 2 - 0.67 & 1.33 & - & 0  & 118 & 346 & 16 \\
            &         & - 1s$^{2}$ 2p $^{2}$P$_{1/2}$  &         &           &       &      &  \\
    \hline
Al X & 637.763 & 2s$^{2}$ $^{1}$S$_{0}$  & 6.15 & E1  & $1.94\times10^{5}$ & 1 - 1.5\tnote{*}  & 1.5 & 0.01 & 1  & 3 & 4.9 & 0.18 \\
            &         & - 2s 2p $^{3}$P$_{1}$  &         &           &       &      &  \\
    \hline
Al X & 670.053 & 2s 2p $^{1}$P$_{1}$  & 6.13 & E1  & $9.38\times10^{8}$ & 1 - 1 & 1 & 106 & 0.35 & 0.38 & 0.61 & 0.07 \\
            &         & – 2p$^{2}$ $^{1}$D$_{2}$  &         &           &       &      &  \\
    \hline
Si IX & 676.503 & 2s$^{2}$ 2p$^{2}$ $^{3}$P$_{1}$  & 6.05 & E1  & $6.54\times10^{4}$ & 1.5 - 2\tnote{*} & 2.25 & $\ll$ 1 & 0.35 & 2.0  & 2.1 & 0.63 \\
            &         & - 2s 2p$^{3}$ $^{5}$S$_{2}$  &         &           &       &       & \\
    \hline
Al IX & 680.318 & 2s$^{2}$ 2p $^{2}$P$_{1/2}$  & 6.02 & E1  & $3.75\times10^{4}$ & 0.67 - 1.73\tnote{*} & 2 & $\ll$ 1 & 0.5  & 4.3 & 4.4 & 0.33 \\
            &         & - 2s 2p$^{2}$ $^{4}$P$_{3/2}$  &         &           &       &      &  \\
    \hline
Na IX & 681.719 & 1s$^{2}$ 2s $^{2}$S$_{1/2}$  & 5.92  & E1  & $6.78\times10^{8}$  & 2 - 1.33 & 1.17 & 58 & 0.5 & 7.6 & 34.6 & 4.9 \\
            &         & - 1s$^{2}$ 2p $^{2}$P$_{3/2}$  &         &           &       &       & \\
    \hline
Mg VIII & 689.641 & 2s 2p$^{2}$ $^{2}$P$_{3/2}$  & \tnote{a}~ 5.95 & E1  & $4.15\times10^{9}$ & 1.33 - 1.2 & 1.1  & 392 & 0.28 & 0.31 &  0.76 & 0.33 \\
            &         & - 2p$^{3}$ $^{2}$D$_{5/2}$  &         &           &       &       & \\
    \hline
Na IX & 694.261 & 1s$^{2}$ 2s $^{2}$S$_{1/2}$  & 5.92 & E1  & $6.42\times10^{8}$  & 2 - 0.67 & 1.33 & - & 0 & 3.5 & 18.4 & 2.6 \\
            &         & - 1s$^{2}$ 2p $^{2}$P$_{1/2}$  &         &           &       & \\
    \hline
\end{tabular}
    \begin{tablenotes}
        \item[a] \citealp{moran2003}
        \item[*]  Land\'e factor of upper and lower energy level has been compared with \cite{verdebout2014}
    \end{tablenotes}
\end{threeparttable}

\end{adjustbox}
\end{turn}
\end{sidewaystable}

\begin{sidewaystable}[htbp]
\begin{adjustbox}{width= 1\textwidth}
\centering

\begin{threeparttable}
\begin{tabular}{llllclllccccc}     
\hline                   
  
\textbf{Ion} & \textbf{$\lambda$(\AA)} & \textbf{Transition (i-k)} &\textbf{log$_{10}$(T)} & \textbf{TT} & \bf{$A(s^{-1})$} &  $\bf{g_{i}-g_{k}}$ & $\bf{\overline{g}}$ & \textbf{B$_{H}$} & \textbf{W$_{2}$}  & \multicolumn{3}{c}{\textbf{L \tiny (ph cm$^{-2}$s$^{-1}$arcsec$^{-2}$)}} \\
  &      &      &   &      &      &   &      &  \textbf{(Gauss)}  &   & \textbf{QS} & \textbf{AR} & \textbf{CH} \\
  \hline
Si IX & 694.686 & 2s$^{2}$ 2p$^{2}$ $^{3}$P$_{2}$  & 6.05 & E1  & $2.16\times10^{5}$ & 1.5 - 2\tnote{*} & 1.75 & 0.01 & 0.35 & 4.1 & 6.7 & 1.5 \\
            &         & - 2s 2p$^{3}$ $^{5}$S$_{2}$  &         &           &       &      &  \\
    \hline
Ar VIII & 700.24 & 2p$^{6}$ 3s $^{2}$S$_{1/2}$  & 5.61 & E1  & $2.7\times10^{9}$  & 2 - 1.33 & 1.17 & 229 & 0.5  & 1.0 & 1.3 & 0.87 \\
            &         & - 2p$^{6}$ 3p $^{2}$P$_{3/2}$ &         &           &       &      &  \\
    \hline
Mg IX & 706.06 & 1s$^{2}$ 2s$^{2}$ $^{1}$S$_{0}$  & 5.99 & E1  & $9.7\times10^{4}$  & 1 - 1.5\tnote{*}  & 1.5 & $\ll$ 1 & 1 & 10.2 & 21.5 & 9.2 \\
            &         & - 1s$^{2}$ 2s 2p $^{3}$P$_{1}$  &         &           &       &      &  \\
   \hline
Ar VIII & 713.801 & 2p$^{6}$ 3s $^{2}$S$_{1/2}$  & 5.61 & E1  & $2.56\times10^{9}$ & 2 - 0.67 & 1.33 & - & 0  & 0.45 & 0.85 & 0.51 \\
            &         & - 2p$^{6}$ 3p $^{2}$P$_{1/2}$  &         &           &       &      &  \\
   \hline
Mg IX & 749.552 & 1s$^{2}$ 2s 2p $^{1}$P$_{1}$  & 5.85–6 & E1  & $7.96\times10^{8}$ & 1 - 1 & 1 & 90 & 0.35  & 1.6 & 4.4 & 1.6 \\
            &         & - 1s$^{2}$ 2p$^{2}$ $^{1}$D$_{2}$ &         &           &       &      &  \\
    \hline
Mg VIII & 762.66 & 2s$^{2}$ 2p $^{2}$P$_{1/2}$  & 5.91 & E1  & $1.82\times10^{4}$ & 0.67 - 1.73\tnote{*} & 2 & $\ll$ 1 & 0.5 & 0.1 & 0.35 & 0.33 \\
            &         & - 2s 2p$^{2}$ $^{4}$P$_{3/2}$  &         &           &       &       & \\
    \hline
Mg VIII & 769.355 & 2s$^{2}$ 2p $^{2}$P$_{1/2}$  & 5.91 & E1  & $1.38\times10^{5}$ & 0.67 - 2.67\tnote{*} & 1.67 & - & 0 & 0.24 & 0.97 & 0.97 \\
            &         & - 2s 2p$^{2}$ $^{4}$P$_{1/2}$  &         &           &       &       & \\
    \hline
Ne VIII & 770.42 & 1s$^{2}$ 2s $^{2}$S$_{1/2}$  & 5.85-6 & E1  & $5.76\times10^{8}$ & 2 - 1.33 & 1.17 & 49 & 0.5 & 39.5 & 177.9 & 41.4 \\
            &         & - 1s$^{2}$ 2p $^{2}$P$_{3/2}$  &         &           &       &       & \\
    \hline
Mg VIII & 772.26 & 2s$^{2}$ 2p $^{2}$P$_{3/2}$  & 5.91 & E1  & $6.06\times10^{4}$ & 1.33 - 1.6\tnote{*}  & 1.8 & $\ll$ 1 & 0.28  & 1.3 & 4.8 & 4.1 \\
            &         & - 2s 2p$^{2}$ $^{4}$P$_{5/2}$  &         &           &       &       & \\
    \hline
S X & 776.373 & 2s$^{2}$2p$^{3}$ $^{4}$S$_{3/2}$  & 6.14 & M1  & $4.92\times10^{2}$  & 2 - 1.33\tnote{*} & 1.66 & $\ll$ 1 & 0.32 & 2.3 & 4.1 & 0.14 \\
            &         & - 2s$^{2}$ 2p$^{3}$ $^{2}$P$_{3/2}$  &         &           &       &       & \\
    \hline
Ne VIII & 780.385 & 1s$^{2}$2s $^{2}$S$_{1/2}$  &  5.85-6 & E1  & $5.54\times10^{8}$ & 2 - 0.67 & 1.33 & - & 0 & 33.1 & 113.9 & 29.7 \\
            &         & - 1s$^{2}$ 2p $^{2}$P$_{1/2}$  &         &           &       &      &  \\
    \hline
Mg VIII & 782.362 & 2s$^{2}$ 2p $^{2}$P$_{3/2}$  & 5.85-6 & E1  & $1.82\times10^{4}$ & 1.33 - 1.73\tnote{*} & 1.53 & $\ll$ 1 & 0.32 & 1.2 & 3.3 & 3.9 \\
            &         & - 2s 2p$^{2}$ $^{4}$P$_{3/2}$  &         &           &       &       & \\
    \hline
S XI & 782.981 & 2s$^{2}$ 2p$^{2}$ $^{3}$P$_{1}$  & 6.15 & M1  & $9.14\times10^{2}$  & 1.5 - 1\tnote{*} & 1.5 & - & 0 & 1.0 & 2.4 & - \\
            &         & - 2s$^{2}$ 2p$^{2}$ $^{1}$S$_{0}$  &         &           &       &       & \\
    \hline
\end{tabular}
    \begin{tablenotes}
        \item[*]  Land\'e factor of upper and lower energy level has been compared with \cite{verdebout2014}
    \end{tablenotes}
\end{threeparttable}
\end{adjustbox}
\end{sidewaystable}

\begin{sidewaystable}[htbp]
\begin{turn}{180}
\centering
\begin{adjustbox}{width= 1\textwidth}

\begin{threeparttable}
\begin{tabular}{llllclllccccc}     
\hline                   
  
\textbf{Ion} & \textbf{$\lambda$(\AA)} & \textbf{Transition (i-k)} &\textbf{log$_{10}$(T)} & \textbf{TT} & \bf{$A(s^{-1})$} &  $\bf{g_{i}-g_{k}}$ & $\bf{\overline{g}}$ & \textbf{B$_{H}$} & \textbf{W$_{2}$} & \multicolumn{3}{c}{\textbf{L \tiny (ph cm$^{-2}$s$^{-1}$arcsec$^{-2}$)}} \\
  &      &      &   &     &      &   &      &  \textbf{(Gauss)}  &  & \textbf{QS} & \textbf{AR} & \textbf{CH} \\
  \hline
S X & 787.557 & 2s$^{2}$ 2p$^{3}$ $^{4}$S$_{3/2}$  & 6.26 & M1  & $1.4\times10^{2}$ & 2 - 0.67\tnote{*} & 2.34 & - & 0  & 1.5 & 2.4 & - \\
            &         & - 2s$^{2}$ 2p$^{3}$ $^{2}$P$_{1/2}$ &         &           &       &       & \\
   \hline
Mg VIII & 789.409 & 2s$^{2}$ 2p $^{2}$P$_{3/2}$  & 5.6 & E1  & $1.38\times10^{5}$ & 1.33 - 2.67\tnote{*} & 0.995 & - & 0  & 0.27 & 1.2 & 0.95 \\
            &         & - 2s 2p$^{2}$ $^{4}$P$_{1/2}$ &         &           &       &       & \\
   \hline
Na VIII & 789.808 & 2s$^{2}$ $^{1}$S$_{0}$  & 5.85-6 & E1  & $4.51\times10^{4}$  & 1 - 1.5\tnote{*}  & 1.5 & $\ll$ 1 & 1  & 0.17 & 1.3 & 0.77 \\
            &         & - 2s 2p $^{3}$P$_{1}$  &         &           &       &       & \\
   \hline
Na VIII & 847.91 & 2s 2p $^{1}$P$_{1}$  & 5.6 & E1  & $6.66\times10^{8}$  & 1 - 1 & 1 & 75 & 0.35 & 0.053 & - & 0.16 \\
            &         & - 2p$^{2}$ $^{1}$D$_{2}$  &         &           &       &       & \\
   \hline
S VIII & 867.88 & 2p$^{4}$ 3s $^{4}$P$_{1/2}$  &  5.85-6 & E1  & $1.56\times10^{9}$ & 2.67 - 1.2  & 0.83 & 147 & 0.5 & 0.068 & - & -  \\
            &         & - 2p$^{4}$ 3p $^{4}$D$_{3/2}$ &         &           &       &       & \\
   \hline
Si IX & 950.083 & 2s$^{2}$ 2p$^{2}$ $^{3}$P$_{1}$  & 5.85-6 & M1  & $2.13\times10^{2}$ & 1.5 - 1\tnote{*} & 1.5 & - & 0  & 2.6 & 5.2 & 1.3 \\
            &         & - 2s$^{2}$ 2p$^{2}$ $^{1}$S$_{0}$  &         &           &       &       & \\
   \hline
O VI & 1031.912 & 2s $^{2}$S$_{1/2}$ & 5.6 & E1  &  $4.16\times10^{8}$  & 2 - 1.33 & 1.17 & 35 & 0.5 & 124 & 260 & 74 \\
           &         & - 2p $^{2}$P$_{3/2}$   &           &            &          &       & \\
    \hline
O VI & 1037.613 & 2s $^{2}$S$_{1/2}$  & 5.6 & E1  & $4.1\times10^{8}$  & 2 - 0.67 & 1.33 & - & 0  & 50.6 & 124 & 33.9 \\
           &        &  - 2p $^{2}$P$_{1/2}$   &          &           &           &      &  \\
    \hline
S X  & 1212.93 & 2s$^{2}$ 2p$^{3}$ $^{4}$S$_{3/2}$  & 6.15 & M1  & $1.56\times10^{1}$ & 2 - 0.8\tnote{*} & 1.4 & $\ll$ 1 & 0.32 & 9.6 & 15.5 & 0.38  \\
            &         & - 2s$^{2}$ 2p$^{3}$ $^{2}$D$_{3/2}$  &         &           &       &      &  \\
   \hline
H I  & 1215.67 & 1s $^{2}$S$_{1/2}$  & 4 - 5.5 & E1  & $6.26\times10^{8}$ & 2 - 1.33 & 1.17 & 53 & 0.5 & \tnote{b}~$1.5\times10^{5}$ & - & -  \\
            &         & - 2p $^{2}$P$_{3/2}$  &         &           &       &      &  \\
   \hline
Fe XII & 1242 & 3s$^{2}$ 3p$^{3}$ $^{4}$S$_{3/2}$  & 6.26 & M1  & $5.4\times10^{2}$  & 2 - 1.33 & 1.67 & $\ll$ 1 & 0.32 & 30.9 & 47.8 & 0.81  \\
            &         & - 3s$^{2}$ 3p$^{3}$ $^{2}$P$_{3/2}$  &         &           &       & \\
  \hline
Fe XII & 1349.4 & 3s$^{2}$ 3p$^{3}$ $^{4}$S$_{3/2}$  & 6.15 & M1  & $2.38\times10^{2}$ & 2 - 0.67 & 2.34 & - & 0 & 18.1 & 28.5 & 0.41 \\
            &         & - 3s$^{2}$ 3p$^{3}$ $^{2}$P$_{1/2}$  &         &           &       &       & \\
   \hline
Fe XI & 1467.07 & 3s$^{2}$ 3p$^{4}$ $^{3}$P$_{1}$  & 6.15 & M1  & $8.87\times10^{2}$ & 1.5 - 1 & 1.5 & - & 0 & 12.5 & 15.5 & 1.5 \\
            &         & - 3s$^{2}$ 3p$^{4}$ $^{1}$S$_{0}$  &         &           &       &      &  \\
   \hline
\end{tabular}
\begin{tablenotes}
    \item[b]  \citealp{Tian2009A&A...504..239T}
    \item[*]  Land\'e factor of upper and lower energy level has been compared with \cite{verdebout2014}
    \end{tablenotes}
\end{threeparttable}

\end{adjustbox}
\end{turn}
\end{sidewaystable}



\begin{acks}[Acknowledgements]
We sincerely thank the anonymous referees for their constructive comments, which helped to improve the overall presentation of the manuscript. We also thank Sarah E. Gibson for her informative comments on the manuscript. Hinode is a Japanese mission developed and launched by ISAS/JAXA, with NAOJ as domestic partner and NASA and STFC (UK) as international partners. It is operated by these agencies in co-operation with ESA and NSC (Norway). The SUMER project was financially supported by DLR, CNES, NASA, and ESA PRODEX Programme (Swiss contribution). SUMER was part of SOHO, the Solar and Heliospheric Observatory, a mission of international cooperation between ESA and NASA. The EUNIS program was supported by the NASA Heliophysics Division through its Low Cost Access to Space Program in Solar and Heliospheric Physics. We acknowledge the use of data of SUMER observations obtained from the Solar Data Analysis Center (SDAC) available at \url{https://sdac.virtualsolar.org}. We also acknowledge the use of spectral lines data of NIST Atomic Spectra Database available at \url{https://physics.nist.gov/PhysRefData/ASD/lines_form.html}.
\end{acks}

\newpage
\bibliographystyle{spr-mp-sola}

\begin{thebibliography}{70}
\ifx\bisbn     \undefined \def\bisbn  #1{ISBN #1}\fi
\ifx\binits    \undefined \def\binits#1{#1}\fi
\ifx\bauthor   \undefined \def\bauthor#1{#1}\fi
\ifx\batitle   \undefined \def\batitle#1{#1}\fi
\ifx\bjtitle   \undefined \def\bjtitle#1{\textit{#1}}\fi
\ifx\bvolume   \undefined \def\bvolume#1{\textbf{#1}}\fi
\ifx\byear     \undefined \def\byear#1{#1}\fi
\ifx\bissue    \undefined \def\bissue#1{#1}\fi
\ifx\bfpage    \undefined \def\bfpage#1{#1}\fi
\ifx\blpage    \undefined \def\blpage #1{#1}\fi
\ifx\burl      \undefined \def\burl#1{\textsf{#1}}\fi
\ifx\href      \undefined \def\href#1#2{\textsf{#2}}\fi
\ifx\betal     \undefined \def\betal{\textit{et al.}}\fi
\ifx\bctitle   \undefined \def\bctitle#1{#1}\fi
\ifx\beditor   \undefined \def\beditor#1{#1}\fi
\ifx\bbtitle   \undefined \def\bbtitle#1{\textit{#1}}\fi
\ifx\bedition  \undefined \def\bedition#1{#1}\fi
\ifx\bseriesno \undefined \def\bseriesno#1{\textbf{#1}}\fi
\ifx\blocation \undefined \def\blocation#1{#1}\fi
\ifx\bsertitle \undefined \def\bsertitle#1{\textit{#1}}\fi
\ifx\bsnm      \undefined \def\bsnm#1{#1}\fi
\ifx\bsuffix   \undefined \def\bsuffix#1{#1}\fi
\ifx\bparticle \undefined \def\bparticle#1{#1}\fi
\ifx\barticle  \undefined \def\barticle#1{}\fi
\ifx\binstitute  \undefined \def\binstitute#1{#1}\fi
\ifx\bpublisher  \undefined \def\bpublisher#1{#1}\fi
\ifx\doiurl    \undefined
  \def\doiurl#1{\href{http://dx.doi.org/#1}{\textsf{DOI}}}\fi
\ifx\arxivurl  \undefined
  \def\arxivurl#1{\href{http://arxiv.org/abs/#1}{\textsf{arXiv}}}\fi
\ifx\adsurl    \undefined
  \def\adsurl#1{\href{http://adsabs.harvard.edu/abs/#1}{\textsf{ADS}}}\fi
\ifx\botherref \undefined \def\botherref#1{}\fi
\ifx\url       \undefined \def\url#1{\textsf{#1}}\fi
\ifx\bchapter  \undefined \def\bchapter#1{}\fi
\ifx\bbook     \undefined \def\bbook#1{}\fi
\ifx\bcomment  \undefined \def\bcomment#1{#1}\fi
\ifx\oauthor   \undefined \def\oauthor#1{#1}\fi
\ifx\citeauthoryear \undefined\def \citeauthoryear#1{#1}\fi
\ifx\endbibitem\undefined \def\endbibitem{}\fi
\ifx\bconflocation  \undefined \def\bconflocation#1{#1} \fi

\bibitem[\protect\citeauthoryear{{Arnaud} and
  {Newkirk}}{1987}]{Arnaud1987A&A...178..263A}
\begin{barticle}
\bauthor{\bsnm{{Arnaud}}, \binits{J.}},
\bauthor{\bsnm{{Newkirk}}, \binits{J.} \bsuffix{G.}}:
\byear{1987},
\batitle{{Mean properties of the polarization of the Fe XIII 10747 A coronal
  emission line}}.
\bjtitle{\aap}
\bvolume{178}(\bissue{1-2}),
\bfpage{263}.
\adsurl{1987A&A...178..263A}.
\end{barticle}
\endbibitem

\bibitem[\protect\citeauthoryear{{B{\c{a}}k-St{\c{e}}{\'s}licka}
  \textit{et~al.}}{2013}]{BS2013ApJ...770L..28B}
\begin{barticle}
\bauthor{\bsnm{{B{\c{a}}k-St{\c{e}}{\'s}licka}}, \binits{U.}},
\bauthor{\bsnm{{Gibson}}, \binits{S.E.}},
\bauthor{\bsnm{{Fan}}, \binits{Y.}},
\bauthor{\bsnm{{Bethge}}, \binits{C.}},
\bauthor{\bsnm{{Forland}}, \binits{B.}},
\bauthor{\bsnm{{Rachmeler}}, \binits{L.A.}}:
\byear{2013},
\batitle{{The Magnetic Structure of Solar Prominence Cavities: New
  Observational Signature Revealed by Coronal Magnetometry}}.
\bjtitle{\apjl}
\bvolume{770}(\bissue{2}),
\bfpage{L28}.
\doiurl{10.1088/2041-8205/770/2/L28}.
\adsurl{2013ApJ...770L..28B}.
\end{barticle}
\endbibitem

\bibitem[\protect\citeauthoryear{{Bommier} and
  {Sahal-Brechot}}{1982}]{bommier1982SoPh...78..157B}
\begin{barticle}
\bauthor{\bsnm{{Bommier}}, \binits{V.}},
\bauthor{\bsnm{{Sahal-Brechot}}, \binits{S.}}:
\byear{1982},
\batitle{{The Hanle Effect of the Coronal L-Alpha Line of Hydrogen -
  Theoretical Investigation}}.
\bjtitle{\solphys}
\bvolume{78}(\bissue{1}),
\bfpage{157}.
\doiurl{10.1007/BF00151151}.
\adsurl{1982SoPh...78..157B}.
\end{barticle}
\endbibitem

\bibitem[\protect\citeauthoryear{{Bommier}, {Leroy}, and
  {Sahal-Br{\'e}chot}}{2021}]{Bommier2021A&A...647A..60B}
\begin{barticle}
\bauthor{\bsnm{{Bommier}}, \binits{V.}},
\bauthor{\bsnm{{Leroy}}, \binits{J.L.}},
\bauthor{\bsnm{{Sahal-Br{\'e}chot}}, \binits{S.}}:
\byear{2021},
\batitle{{24 synoptic maps of average magnetic field in 296 prominences
  measured by the Hanle effect during the ascending phase of solar cycle 21}}.
\bjtitle{\aap}
\bvolume{647},
\bfpage{A60}.
\doiurl{10.1051/0004-6361/202038868}.
\adsurl{2021A&A...647A..60B}.
\end{barticle}
\endbibitem

\bibitem[\protect\citeauthoryear{{Bommier}, {Sahal-Brechot}, and
  {Leroy}}{1981}]{Bommier1981A&A...100..231B}
\begin{barticle}
\bauthor{\bsnm{{Bommier}}, \binits{V.}},
\bauthor{\bsnm{{Sahal-Brechot}}, \binits{S.}},
\bauthor{\bsnm{{Leroy}}, \binits{J.L.}}:
\byear{1981},
\batitle{{Determination of the complete vector magnetic field in solar
  prominences, using the Hanle effect}}.
\bjtitle{\aap}
\bvolume{100}(\bissue{2}),
\bfpage{231}.
\adsurl{1981A&A...100..231B}.
\end{barticle}
\endbibitem

\bibitem[\protect\citeauthoryear{{Bommier}
  \textit{et~al.}}{1994}]{Bommier1994SoPh..154..231B}
\begin{barticle}
\bauthor{\bsnm{{Bommier}}, \binits{V.}},
\bauthor{\bsnm{{Landi Degl'Innocenti}}, \binits{E.}},
\bauthor{\bsnm{{Leroy}}, \binits{J.-L.}},
\bauthor{\bsnm{{Sahal-Brechot}}, \binits{S.}}:
\byear{1994},
\batitle{{Complete determination of the magnetic field vector and of the
  electron density in 14 prominences from linear polarizaton measurements in
  the HeI D$_{3}$ and H{\ensuremath{\alpha}} lines}}.
\bjtitle{\solphys}
\bvolume{154}(\bissue{2}),
\bfpage{231}.
\doiurl{10.1007/BF00681098}.
\adsurl{1994SoPh..154..231B}.
\end{barticle}
\endbibitem

\bibitem[\protect\citeauthoryear{{Brosius}, {Rabin}, and
  {Thomas}}{2007}]{Brosius2007ApJ...656L..41B}
\begin{barticle}
\bauthor{\bsnm{{Brosius}}, \binits{J.W.}},
\bauthor{\bsnm{{Rabin}}, \binits{D.M.}},
\bauthor{\bsnm{{Thomas}}, \binits{R.J.}}:
\byear{2007},
\batitle{{Doppler Velocities Measured in Coronal Emission Lines from a Bright
  Point Observed with the EUNIS Sounding Rocket}}.
\bjtitle{\apjl}
\bvolume{656}(\bissue{1}),
\bfpage{L41}.
\doiurl{10.1086/512185}.
\adsurl{2007ApJ...656L..41B}.
\end{barticle}
\endbibitem

\bibitem[\protect\citeauthoryear{{Brosius}
  \textit{et~al.}}{2008}]{Brosius2008ApJ...677..781B}
\begin{barticle}
\bauthor{\bsnm{{Brosius}}, \binits{J.W.}},
\bauthor{\bsnm{{Rabin}}, \binits{D.M.}},
\bauthor{\bsnm{{Thomas}}, \binits{R.J.}},
\bauthor{\bsnm{{Landi}}, \binits{E.}}:
\byear{2008},
\batitle{{Analysis of a Solar Coronal Bright Point Extreme Ultraviolet Spectrum
  from the EUNIS Sounding Rocket Instrument}}.
\bjtitle{\apj}
\bvolume{677}(\bissue{1}),
\bfpage{781}.
\doiurl{10.1086/528930}.
\adsurl{2008ApJ...677..781B}.
\end{barticle}
\endbibitem

\bibitem[\protect\citeauthoryear{{Casini} and
  {Judge}}{1999}]{Casini1999ApJ...522..524C}
\begin{barticle}
\bauthor{\bsnm{{Casini}}, \binits{R.}},
\bauthor{\bsnm{{Judge}}, \binits{P.G.}}:
\byear{1999},
\batitle{{Spectral Lines for Polarization Measurements of the Coronal Magnetic
  Field. II. Consistent Treatment of the Stokes Vector forMagnetic-Dipole
  Transitions}}.
\bjtitle{\apj}
\bvolume{522}(\bissue{1}),
\bfpage{524}.
\doiurl{10.1086/307629}.
\adsurl{1999ApJ...522..524C}.
\end{barticle}
\endbibitem

\bibitem[\protect\citeauthoryear{{Condon} and
  {Shortley}}{1935}]{Condon1935tas..book.....C}
\begin{bbook}
\bauthor{\bsnm{{Condon}}, \binits{E.U.}},
\bauthor{\bsnm{{Shortley}}, \binits{G.H.}}:
\byear{1935},
\bbtitle{{The Theory of Atomic Spectra}}.
\adsurl{1935tas..book.....C}.
\end{bbook}
\endbibitem

\bibitem[\protect\citeauthoryear{{Culhane}}{2007}]{Culhane2007ASPC..369....3C}
\begin{bchapter}
\bauthor{\bsnm{{Culhane}}, \binits{J.L.}}:
\byear{2007},
\bctitle{{The Solar-B EUV Imaging Spectrometer: an Overview of the EIS
  Instrument}}.
In: \beditor{\bsnm{{Shibata}}, \binits{K.}},
\beditor{\bsnm{{Nagata}}, \binits{S.}},
\beditor{\bsnm{{Sakurai}}, \binits{T.}} (eds.)
\bbtitle{New Solar Physics with Solar-B Mission},
\bsertitle{Astronomical Society of the Pacific Conference Series}
\bseriesno{369},
\bfpage{3}.
\adsurl{2007ASPC..369....3C}.
\end{bchapter}
\endbibitem

\bibitem[\protect\citeauthoryear{{Curdt} and {Landi}}{2001}]{curdt2001spectral}
\begin{bchapter}
\bauthor{\bsnm{{Curdt}}, \binits{W.}},
\bauthor{\bsnm{{Landi}}, \binits{E.}}:
\byear{2001},
\bctitle{{Spectral windows of the solar atmosphere}}.
In: \beditor{\bsnm{{Battrick}}, \binits{B.}},
\beditor{\bsnm{{Sawaya-Lacoste}}, \binits{H.}},
\beditor{\bsnm{{Marsch}}, \binits{E.}},
\beditor{\bsnm{{Martinez Pillet}}, \binits{V.}},
\beditor{\bsnm{{Fleck}}, \binits{B.}},
\beditor{\bsnm{{Marsden}}, \binits{R.}} (eds.)
\bbtitle{Solar encounter. Proceedings of the First Solar Orbiter Workshop},
\bsertitle{ESA Special Publication}
\bseriesno{493},
\bfpage{199}.
\adsurl{2001ESASP.493..199C}.
\end{bchapter}
\endbibitem

\bibitem[\protect\citeauthoryear{{Curdt}, {Landi}, and
  {Feldman}}{2004}]{curdt2004sumer}
\begin{barticle}
\bauthor{\bsnm{{Curdt}}, \binits{W.}},
\bauthor{\bsnm{{Landi}}, \binits{E.}},
\bauthor{\bsnm{{Feldman}}, \binits{U.}}:
\byear{2004},
\batitle{{The SUMER spectral atlas of solar coronal features}}.
\bjtitle{\aap}
\bvolume{427},
\bfpage{1045}.
\doiurl{10.1051/0004-6361:20041278}.
\adsurl{2004A&A...427.1045C}.
\end{barticle}
\endbibitem

\bibitem[\protect\citeauthoryear{{Del Zanna}}{2012}]{Zanna2012A&A...537A..38D}
\begin{barticle}
\bauthor{\bsnm{{Del Zanna}}, \binits{G.}}:
\byear{2012},
\batitle{{Benchmarking atomic data for the CHIANTI atomic database: coronal
  lines observed by Hinode EIS}}.
\bjtitle{\aap}
\bvolume{537},
\bfpage{A38}.
\doiurl{10.1051/0004-6361/201117592}.
\adsurl{2012A&A...537A..38D}.
\end{barticle}
\endbibitem

\bibitem[\protect\citeauthoryear{{Del Zanna}
  \textit{et~al.}}{2021}]{DelZanna2021ApJ...909...38D}
\begin{barticle}
\bauthor{\bsnm{{Del Zanna}}, \binits{G.}},
\bauthor{\bsnm{{Dere}}, \binits{K.P.}},
\bauthor{\bsnm{{Young}}, \binits{P.R.}},
\bauthor{\bsnm{{Landi}}, \binits{E.}}:
\byear{2021},
\batitle{{CHIANTI{\textemdash}An Atomic Database for Emission Lines. XVI.
  Version 10, Further Extensions}}.
\bjtitle{\apj}
\bvolume{909}(\bissue{1}),
\bfpage{38}.
\doiurl{10.3847/1538-4357/abd8ce}.
\adsurl{2021ApJ...909...38D}.
\end{barticle}
\endbibitem

\bibitem[\protect\citeauthoryear{{Dere}
  \textit{et~al.}}{1997}]{Dere1997A&AS..125..149D}
\begin{barticle}
\bauthor{\bsnm{{Dere}}, \binits{K.P.}},
\bauthor{\bsnm{{Landi}}, \binits{E.}},
\bauthor{\bsnm{{Mason}}, \binits{H.E.}},
\bauthor{\bsnm{{Monsignori Fossi}}, \binits{B.C.}},
\bauthor{\bsnm{{Young}}, \binits{P.R.}}:
\byear{1997},
\batitle{{CHIANTI - an atomic database for emission lines}}.
\bjtitle{\aaps}
\bvolume{125},
\bfpage{149}.
\doiurl{10.1051/aas:1997368}.
\adsurl{1997A&AS..125..149D}.
\end{barticle}
\endbibitem

\bibitem[\protect\citeauthoryear{{Drake}}{2006}]{Drake2006sham.book.....D}
\begin{bbook}
\bauthor{\bsnm{{Drake}}, \binits{G.W.F.}}:
\byear{2006},
\bbtitle{{Springer Handbook of Atomic, Molecular, and Optical Physics}}.
\doiurl{10.1007/978-0-387-26308-3}.
\adsurl{2006sham.book.....D}.
\end{bbook}
\endbibitem

\bibitem[\protect\citeauthoryear{{Feldman} \textit{et~al.}}{1997}]{feldman1997}
\begin{barticle}
\bauthor{\bsnm{{Feldman}}, \binits{U.}},
\bauthor{\bsnm{{Behring}}, \binits{W.E.}},
\bauthor{\bsnm{{Curdt}}, \binits{W.}},
\bauthor{\bsnm{{Sch{\"u}hle}}, \binits{U.}},
\bauthor{\bsnm{{Wilhelm}}, \binits{K.}},
\bauthor{\bsnm{{Lemaire}}, \binits{P.}},
\bauthor{\bsnm{{Moran}}, \binits{T.M.}}:
\byear{1997},
\batitle{{A Coronal Spectrum in the 500--1610 Angstrom Wavelength Range
  Recorded at a Height of 21,000 Kilometers above the West Solar Limb by the
  SUMER Instrument on Solar and Heliospheric Observatory}}.
\bjtitle{\apjs}
\bvolume{113}(\bissue{1}),
\bfpage{195}.
\doiurl{10.1086/313048}.
\adsurl{1997ApJS..113..195F}.
\end{barticle}
\endbibitem

\bibitem[\protect\citeauthoryear{{Feldman}
  \textit{et~al.}}{1998}]{Feldman1998ApJ...502..997F}
\begin{barticle}
\bauthor{\bsnm{{Feldman}}, \binits{U.}},
\bauthor{\bsnm{{Brown}}, \binits{C.M.}},
\bauthor{\bsnm{{Laming}}, \binits{J.M.}},
\bauthor{\bsnm{{Seely}}, \binits{J.F.}},
\bauthor{\bsnm{{Doschek}}, \binits{G.A.}}:
\byear{1998},
\batitle{{A Compact Spectral Range and Matching Extreme-Ultraviolet
  Spectrometer for the Simultaneous Study of 1 {\texttimes} {}10$^{4}$-2
  {\texttimes} {}10$^{7}$ K Solar Plasmas}}.
\bjtitle{\apj}
\bvolume{502}(\bissue{2}),
\bfpage{997}.
\doiurl{10.1086/305931}.
\adsurl{1998ApJ...502..997F}.
\end{barticle}
\endbibitem

\bibitem[\protect\citeauthoryear{{Fineschi}
  \textit{et~al.}}{1993}]{Fineschi1993SPIE.1742..423F}
\begin{bchapter}
\bauthor{\bsnm{{Fineschi}}, \binits{S.}},
\bauthor{\bsnm{{Hoover}}, \binits{R.B.}},
\bauthor{\bsnm{{Zukic}}, \binits{M.}},
\bauthor{\bsnm{{Kim}}, \binits{J.}},
\bauthor{\bsnm{{Walker}}, \binits{J.} \bsuffix{Arthur B.~C.}},
\bauthor{\bsnm{{Baker}}, \binits{P.C.}}:
\byear{1993},
\bctitle{{Polarimetry of HI Lyman-alpha for coronal magnetic field
  diagnostics}}.
In: \beditor{\bsnm{{Hoover}}, \binits{R.B.}},
\beditor{\bsnm{{Walker}}, \binits{J.} \bsuffix{Arthur B.~C.}} (eds.)
\bbtitle{Multilayer and Grazing Incidence X-Ray/EUV Optics for Astronomy and
  Projection Lithography},
\bsertitle{Society of Photo-Optical Instrumentation Engineers (SPIE) Conference
  Series}
\bseriesno{1742},
\bfpage{423}.
\doiurl{10.1117/12.140576}.
\adsurl{1993SPIE.1742..423F}.
\end{bchapter}
\endbibitem

\bibitem[\protect\citeauthoryear{{Gibson}
  \textit{et~al.}}{2017}]{Gibson2017ApJ...840L..13G}
\begin{barticle}
\bauthor{\bsnm{{Gibson}}, \binits{S.E.}},
\bauthor{\bsnm{{Dalmasse}}, \binits{K.}},
\bauthor{\bsnm{{Rachmeler}}, \binits{L.A.}},
\bauthor{\bsnm{{De Rosa}}, \binits{M.L.}},
\bauthor{\bsnm{{Tomczyk}}, \binits{S.}},
\bauthor{\bsnm{{de Toma}}, \binits{G.}},
\bauthor{\bsnm{{Burkepile}}, \binits{J.}},
\bauthor{\bsnm{{Galloy}}, \binits{M.}}:
\byear{2017},
\batitle{{Magnetic Nulls and Super-radial Expansion in the Solar Corona}}.
\bjtitle{\apjl}
\bvolume{840}(\bissue{2}),
\bfpage{L13}.
\doiurl{10.3847/2041-8213/aa6fac}.
\adsurl{2017ApJ...840L..13G}.
\end{barticle}
\endbibitem

\bibitem[\protect\citeauthoryear{{Harvey}}{1969}]{Harvey1969PhDT.........3H}
\begin{botherref}
\oauthor{\bsnm{{Harvey}}, \binits{J.W.}}:
1969,
{Magnetic Fields Associated with Solar Active-Region Prominences.}
PhD thesis,
National Solar Observatory.
\adsurl{1969PhDT.........3H}.
\end{botherref}
\endbibitem

\bibitem[\protect\citeauthoryear{{Hebbur Dayananda}
  \textit{et~al.}}{2021}]{Hebbur2021ApJ...920..140H}
\begin{barticle}
\bauthor{\bsnm{{Hebbur Dayananda}}, \binits{S.}},
\bauthor{\bsnm{{Trujillo Bueno}}, \binits{J.}},
\bauthor{\bsnm{{de Vicente}}, \binits{{\'A}.}},
\bauthor{\bsnm{{del Pino Alem{\'a}n}}, \binits{T.}}:
\byear{2021},
\batitle{{Polarization of the Ly{\ensuremath{\alpha}} Lines of H I and He II as
  a Tool for Exploring the Solar Corona}}.
\bjtitle{\apj}
\bvolume{920}(\bissue{2}),
\bfpage{140}.
\doiurl{10.3847/1538-4357/ac1068}.
\adsurl{2021ApJ...920..140H}.
\end{barticle}
\endbibitem

\bibitem[\protect\citeauthoryear{{Ishikawa}
  \textit{et~al.}}{2021}]{Ishikawa2021SciA....7.8406I}
\begin{barticle}
\bauthor{\bsnm{{Ishikawa}}, \binits{R.}},
\bauthor{\bsnm{{Bueno}}, \binits{J.T.}},
\bauthor{\bsnm{{del Pino Alem{\'a}n}}, \binits{T.}},
\bauthor{\bsnm{{Okamoto}}, \binits{T.J.}},
\bauthor{\bsnm{{McKenzie}}, \binits{D.E.}},
\bauthor{\bsnm{{Auch{\`e}re}}, \binits{F.}},
\bauthor{\bsnm{{Kano}}, \binits{R.}},
\bauthor{\bsnm{{Song}}, \binits{D.}},
\bauthor{\bsnm{{Yoshida}}, \binits{M.}},
\bauthor{\bsnm{{Rachmeler}}, \binits{L.A.}},
\bauthor{\bsnm{{Kobayashi}}, \binits{K.}},
\bauthor{\bsnm{{Hara}}, \binits{H.}},
\bauthor{\bsnm{{Kubo}}, \binits{M.}},
\bauthor{\bsnm{{Narukage}}, \binits{N.}},
\bauthor{\bsnm{{Sakao}}, \binits{T.}},
\bauthor{\bsnm{{Shimizu}}, \binits{T.}},
\bauthor{\bsnm{{Suematsu}}, \binits{Y.}},
\bauthor{\bsnm{{Bethge}}, \binits{C.}},
\bauthor{\bsnm{{De Pontieu}}, \binits{B.}},
\bauthor{\bsnm{{Dalda}}, \binits{A.S.}},
\bauthor{\bsnm{{Vigil}}, \binits{G.D.}},
\bauthor{\bsnm{{Winebarger}}, \binits{A.}},
\bauthor{\bsnm{{Ballester}}, \binits{E.A.}},
\bauthor{\bsnm{{Belluzzi}}, \binits{L.}},
\bauthor{\bsnm{{{\v{S}}t{\v{e}}p{\'a}n}}, \binits{J.}},
\bauthor{\bsnm{{Ramos}}, \binits{A.A.}},
\bauthor{\bsnm{{Carlsson}}, \binits{M.}},
\bauthor{\bsnm{{Leenaarts}}, \binits{J.}}:
\byear{2021},
\batitle{{Mapping solar magnetic fields from the photosphere to the base of the
  corona}}.
\bjtitle{Science Advances}
\bvolume{7}(\bissue{8}),
\bfpage{eabe8406}.
\doiurl{10.1126/sciadv.abe8406}.
\adsurl{2021SciA....7.8406I}.
\end{barticle}
\endbibitem

\bibitem[\protect\citeauthoryear{{J{\"o}nsson}
  \textit{et~al.}}{2007}]{Jonsson2007CoPhC.177..597J}
\begin{barticle}
\bauthor{\bsnm{{J{\"o}nsson}}, \binits{P.}},
\bauthor{\bsnm{{He}}, \binits{X.}},
\bauthor{\bsnm{{Froese Fischer}}, \binits{C.}},
\bauthor{\bsnm{{Grant}}, \binits{I.P.}}:
\byear{2007},
\batitle{{The grasp2K relativistic atomic structure package}}.
\bjtitle{Computer Physics Communications}
\bvolume{177}(\bissue{7}),
\bfpage{597}.
\doiurl{10.1016/j.cpc.2007.06.002}.
\adsurl{2007CoPhC.177..597J}.
\end{barticle}
\endbibitem

\bibitem[\protect\citeauthoryear{{Judge}}{1998}]{Judge1998ApJ...500.1009J}
\begin{barticle}
\bauthor{\bsnm{{Judge}}, \binits{P.G.}}:
\byear{1998},
\batitle{{Spectral Lines for Polarization Measurements of the Coronal Magnetic
  Field. I. Theoretical Intensities}}.
\bjtitle{\apj}
\bvolume{500}(\bissue{2}),
\bfpage{1009}.
\doiurl{10.1086/305775}.
\adsurl{1998ApJ...500.1009J}.
\end{barticle}
\endbibitem

\bibitem[\protect\citeauthoryear{{Kano}
  \textit{et~al.}}{2017}]{Kano2017ApJ...839L..10K}
\begin{barticle}
\bauthor{\bsnm{{Kano}}, \binits{R.}},
\bauthor{\bsnm{{Trujillo Bueno}}, \binits{J.}},
\bauthor{\bsnm{{Winebarger}}, \binits{A.}},
\bauthor{\bsnm{{Auch{\`e}re}}, \binits{F.}},
\bauthor{\bsnm{{Narukage}}, \binits{N.}},
\bauthor{\bsnm{{Ishikawa}}, \binits{R.}},
\bauthor{\bsnm{{Kobayashi}}, \binits{K.}},
\bauthor{\bsnm{{Bando}}, \binits{T.}},
\bauthor{\bsnm{{Katsukawa}}, \binits{Y.}},
\bauthor{\bsnm{{Kubo}}, \binits{M.}},
\bauthor{\bsnm{{Ishikawa}}, \binits{S.}},
\bauthor{\bsnm{{Giono}}, \binits{G.}},
\bauthor{\bsnm{{Hara}}, \binits{H.}},
\bauthor{\bsnm{{Suematsu}}, \binits{Y.}},
\bauthor{\bsnm{{Shimizu}}, \binits{T.}},
\bauthor{\bsnm{{Sakao}}, \binits{T.}},
\bauthor{\bsnm{{Tsuneta}}, \binits{S.}},
\bauthor{\bsnm{{Ichimoto}}, \binits{K.}},
\bauthor{\bsnm{{Goto}}, \binits{M.}},
\bauthor{\bsnm{{Belluzzi}}, \binits{L.}},
\bauthor{\bsnm{{{\v{S}}t{\v{e}}p{\'a}n}}, \binits{J.}},
\bauthor{\bsnm{{Asensio Ramos}}, \binits{A.}},
\bauthor{\bsnm{{Manso Sainz}}, \binits{R.}},
\bauthor{\bsnm{{Champey}}, \binits{P.}},
\bauthor{\bsnm{{Cirtain}}, \binits{J.}},
\bauthor{\bsnm{{De Pontieu}}, \binits{B.}},
\bauthor{\bsnm{{Casini}}, \binits{R.}},
\bauthor{\bsnm{{Carlsson}}, \binits{M.}}:
\byear{2017},
\batitle{{Discovery of Scattering Polarization in the Hydrogen
  Ly{\ensuremath{\alpha}} Line of the Solar Disk Radiation}}.
\bjtitle{\apjl}
\bvolume{839}(\bissue{1}),
\bfpage{L10}.
\doiurl{10.3847/2041-8213/aa697f}.
\adsurl{2017ApJ...839L..10K}.
\end{barticle}
\endbibitem

\bibitem[\protect\citeauthoryear{{Kano}
  \textit{et~al.}}{2019}]{Kano2019AAS...23430216K}
\begin{bchapter}
\bauthor{\bsnm{{Kano}}, \binits{R.}},
\bauthor{\bsnm{{Ishikawa}}, \binits{R.}},
\bauthor{\bsnm{{McKenzie}}, \binits{D.E.}},
\bauthor{\bsnm{{Trujillo Bueno}}, \binits{J.}},
\bauthor{\bsnm{{Song}}, \binits{D.}},
\bauthor{\bsnm{{Yoshida}}, \binits{M.}},
\bauthor{\bsnm{{Okamoto}}, \binits{T.}},
\bauthor{\bsnm{{Rachmeler}}, \binits{L.}},
\bauthor{\bsnm{{Kobayashi}}, \binits{K.}},
\bauthor{\bsnm{{Auchere}}, \binits{F.}}:
\byear{2019},
\bctitle{{Lyman-{\ensuremath{\alpha}} imaging polarimetry with the CLASP2
  sounding rocket mission}}.
In: \bbtitle{American Astronomical Society Meeting Abstracts \#234},
\bsertitle{American Astronomical Society Meeting Abstracts}
\bseriesno{234},
\bfpage{302.16}.
\adsurl{2019AAS...23430216K}.
\end{bchapter}
\endbibitem

\bibitem[\protect\citeauthoryear{{Kishore}
  \textit{et~al.}}{2015}]{Kishore2015SoPh..290.2409K}
\begin{barticle}
\bauthor{\bsnm{{Kishore}}, \binits{P.}},
\bauthor{\bsnm{{Ramesh}}, \binits{R.}},
\bauthor{\bsnm{{Kathiravan}}, \binits{C.}},
\bauthor{\bsnm{{Rajalingam}}, \binits{M.}}:
\byear{2015},
\batitle{{A Low-Frequency Radio Spectropolarimeter for Observations of the
  Solar Corona}}.
\bjtitle{\solphys}
\bvolume{290}(\bissue{9}),
\bfpage{2409}.
\doiurl{10.1007/s11207-015-0705-0}.
\adsurl{2015SoPh..290.2409K}.
\end{barticle}
\endbibitem

\bibitem[\protect\citeauthoryear{{Kumari}
  \textit{et~al.}}{2019}]{Kumari2019ApJ...881...24K}
\begin{barticle}
\bauthor{\bsnm{{Kumari}}, \binits{A.}},
\bauthor{\bsnm{{Ramesh}}, \binits{R.}},
\bauthor{\bsnm{{Kathiravan}}, \binits{C.}},
\bauthor{\bsnm{{Wang}}, \binits{T.J.}},
\bauthor{\bsnm{{Gopalswamy}}, \binits{N.}}:
\byear{2019},
\batitle{{Direct Estimates of the Solar Coronal Magnetic Field Using
  Contemporaneous Extreme-ultraviolet, Radio, and White-light Observations}}.
\bjtitle{\apj}
\bvolume{881}(\bissue{1}),
\bfpage{24}.
\doiurl{10.3847/1538-4357/ab2adf}.
\adsurl{2019ApJ...881...24K}.
\end{barticle}
\endbibitem

\bibitem[\protect\citeauthoryear{{Lagg}
  \textit{et~al.}}{2017}]{Lagg2017SSRv..210...37L}
\begin{barticle}
\bauthor{\bsnm{{Lagg}}, \binits{A.}},
\bauthor{\bsnm{{Lites}}, \binits{B.}},
\bauthor{\bsnm{{Harvey}}, \binits{J.}},
\bauthor{\bsnm{{Gosain}}, \binits{S.}},
\bauthor{\bsnm{{Centeno}}, \binits{R.}}:
\byear{2017},
\batitle{{Measurements of Photospheric and Chromospheric Magnetic Fields}}.
\bjtitle{\ssr}
\bvolume{210}(\bissue{1-4}),
\bfpage{37}.
\doiurl{10.1007/s11214-015-0219-y}.
\adsurl{2017SSRv..210...37L}.
\end{barticle}
\endbibitem

\bibitem[\protect\citeauthoryear{{Landi}
  \textit{et~al.}}{2020}]{Landi2020ApJ...904...87L}
\begin{barticle}
\bauthor{\bsnm{{Landi}}, \binits{E.}},
\bauthor{\bsnm{{Hutton}}, \binits{R.}},
\bauthor{\bsnm{{Brage}}, \binits{T.}},
\bauthor{\bsnm{{Li}}, \binits{W.}}:
\byear{2020},
\batitle{{Hinode/EIS Measurements of Active-region Magnetic Fields}}.
\bjtitle{\apj}
\bvolume{904}(\bissue{2}),
\bfpage{87}.
\doiurl{10.3847/1538-4357/abbf54}.
\adsurl{2020ApJ...904...87L}.
\end{barticle}
\endbibitem

\bibitem[\protect\citeauthoryear{{Landi}
  \textit{et~al.}}{2021}]{Landi2021ApJ...913....1L}
\begin{barticle}
\bauthor{\bsnm{{Landi}}, \binits{E.}},
\bauthor{\bsnm{{Li}}, \binits{W.}},
\bauthor{\bsnm{{Brage}}, \binits{T.}},
\bauthor{\bsnm{{Hutton}}, \binits{R.}}:
\byear{2021},
\batitle{{Hinode/EIS Coronal Magnetic Field Measurements at the Onset of a C2
  Flare}}.
\bjtitle{\apj}
\bvolume{913}(\bissue{1}),
\bfpage{1}.
\doiurl{10.3847/1538-4357/abf6d1}.
\adsurl{2021ApJ...913....1L}.
\end{barticle}
\endbibitem

\bibitem[\protect\citeauthoryear{{Landi
  Degl'Innocenti}}{1982}]{LandiDeglInnocenti1982SoPh...77..285L}
\begin{barticle}
\bauthor{\bsnm{{Landi Degl'Innocenti}}, \binits{E.}}:
\byear{1982},
\batitle{{On the effective Land{\'e} factor of magnetic lines}}.
\bjtitle{\solphys}
\bvolume{77}(\bissue{1-2}),
\bfpage{285}.
\doiurl{10.1007/BF00156111}.
\adsurl{1982SoPh...77..285L}.
\end{barticle}
\endbibitem

\bibitem[\protect\citeauthoryear{{Landi Degl'Innocenti} and
  {Landolfi}}{2004}]{Landi2004ASSL..307.....L}
\begin{bbook}
\bauthor{\bsnm{{Landi Degl'Innocenti}}, \binits{E.}},
\bauthor{\bsnm{{Landolfi}}, \binits{M.}}:
\byear{2004},
\bbtitle{{Polarization in Spectral Lines}}
\bseriesno{307}.
\doiurl{10.1007/BF00156111}.
\adsurl{2004ASSL..307.....L}.
\end{bbook}
\endbibitem

\bibitem[\protect\citeauthoryear{{Li}
  \textit{et~al.}}{2016}]{Li2016ApJ...826..219L}
\begin{barticle}
\bauthor{\bsnm{{Li}}, \binits{W.}},
\bauthor{\bsnm{{Yang}}, \binits{Y.}},
\bauthor{\bsnm{{Tu}}, \binits{B.}},
\bauthor{\bsnm{{Xiao}}, \binits{J.}},
\bauthor{\bsnm{{Grumer}}, \binits{J.}},
\bauthor{\bsnm{{Brage}}, \binits{T.}},
\bauthor{\bsnm{{Watanabe}}, \binits{T.}},
\bauthor{\bsnm{{Hutton}}, \binits{R.}},
\bauthor{\bsnm{{Zou}}, \binits{Y.}}:
\byear{2016},
\batitle{{Atomic-level Pseudo-degeneracy of Atomic Levels Giving Transitions
  Induced by Magnetic Fields, of Importance for Determining the Field Strengths
  in the Solar Corona}}.
\bjtitle{\apj}
\bvolume{826}(\bissue{2}),
\bfpage{219}.
\doiurl{10.3847/0004-637X/826/2/219}.
\adsurl{2016ApJ...826..219L}.
\end{barticle}
\endbibitem

\bibitem[\protect\citeauthoryear{{Li}
  \textit{et~al.}}{2021}]{Li2021ApJ...913..135L}
\begin{barticle}
\bauthor{\bsnm{{Li}}, \binits{W.}},
\bauthor{\bsnm{{Li}}, \binits{M.}},
\bauthor{\bsnm{{Wang}}, \binits{K.}},
\bauthor{\bsnm{{Brage}}, \binits{T.}},
\bauthor{\bsnm{{Hutton}}, \binits{R.}},
\bauthor{\bsnm{{Landi}}, \binits{E.}}:
\byear{2021},
\batitle{{A Theoretical Investigation of the Magnetic-field-induced Transition
  in Fe X, of Importance for Measuring Magnetic Field Strengths in the Solar
  Corona}}.
\bjtitle{\apj}
\bvolume{913}(\bissue{2}),
\bfpage{135}.
\doiurl{10.3847/1538-4357/abfa97}.
\adsurl{2021ApJ...913..135L}.
\end{barticle}
\endbibitem

\bibitem[\protect\citeauthoryear{{Lin}, {Kuhn}, and
  {Coulter}}{2004}]{Lin2004ApJ...613L.177L}
\begin{barticle}
\bauthor{\bsnm{{Lin}}, \binits{H.}},
\bauthor{\bsnm{{Kuhn}}, \binits{J.R.}},
\bauthor{\bsnm{{Coulter}}, \binits{R.}}:
\byear{2004},
\batitle{{Coronal Magnetic Field Measurements}}.
\bjtitle{\apjl}
\bvolume{613}(\bissue{2}),
\bfpage{L177}.
\doiurl{10.1086/425217}.
\adsurl{2004ApJ...613L.177L}.
\end{barticle}
\endbibitem

\bibitem[\protect\citeauthoryear{{Lin}, {Penn}, and
  {Tomczyk}}{2000}]{Lin2000ApJ...541L..83L}
\begin{barticle}
\bauthor{\bsnm{{Lin}}, \binits{H.}},
\bauthor{\bsnm{{Penn}}, \binits{M.J.}},
\bauthor{\bsnm{{Tomczyk}}, \binits{S.}}:
\byear{2000},
\batitle{{A New Precise Measurement of the Coronal Magnetic Field Strength}}.
\bjtitle{\apjl}
\bvolume{541}(\bissue{2}),
\bfpage{L83}.
\doiurl{10.1086/312900}.
\adsurl{2000ApJ...541L..83L}.
\end{barticle}
\endbibitem

\bibitem[\protect\citeauthoryear{{McIntosh}
  \textit{et~al.}}{2011}]{McIntosh2011Natur.475..477M}
\begin{barticle}
\bauthor{\bsnm{{McIntosh}}, \binits{S.W.}},
\bauthor{\bsnm{{de Pontieu}}, \binits{B.}},
\bauthor{\bsnm{{Carlsson}}, \binits{M.}},
\bauthor{\bsnm{{Hansteen}}, \binits{V.}},
\bauthor{\bsnm{{Boerner}}, \binits{P.}},
\bauthor{\bsnm{{Goossens}}, \binits{M.}}:
\byear{2011},
\batitle{{Alfv{\'e}nic waves with sufficient energy to power the quiet solar
  corona and fast solar wind}}.
\bjtitle{\nat}
\bvolume{475}(\bissue{7357}),
\bfpage{477}.
\doiurl{10.1038/nature10235}.
\adsurl{2011Natur.475..477M}.
\end{barticle}
\endbibitem

\bibitem[\protect\citeauthoryear{{Moran}}{2003}]{moran2003}
\begin{barticle}
\bauthor{\bsnm{{Moran}}, \binits{T.G.}}:
\byear{2003},
\batitle{{Test for Alfv{\'e}n Wave Signatures in a Solar Coronal Hole}}.
\bjtitle{\apj}
\bvolume{598}(\bissue{1}),
\bfpage{657}.
\doiurl{10.1086/378795}.
\adsurl{2003ApJ...598..657M}.
\end{barticle}
\endbibitem

\bibitem[\protect\citeauthoryear{{Mugundhan}
  \textit{et~al.}}{2018}]{Mugundhan2018SoPh..293...41M}
\begin{barticle}
\bauthor{\bsnm{{Mugundhan}}, \binits{V.}},
\bauthor{\bsnm{{Ramesh}}, \binits{R.}},
\bauthor{\bsnm{{Kathiravan}}, \binits{C.}},
\bauthor{\bsnm{{Gireesh}}, \binits{G.V.S.}},
\bauthor{\bsnm{{Hegde}}, \binits{A.}}:
\byear{2018},
\batitle{{Spectropolarimetric Observations of Solar Noise Storms at Low
  Frequencies}}.
\bjtitle{\solphys}
\bvolume{293}(\bissue{3}),
\bfpage{41}.
\doiurl{10.1007/s11207-018-1260-2}.
\adsurl{2018SoPh..293...41M}.
\end{barticle}
\endbibitem

\bibitem[\protect\citeauthoryear{{Nagaraju}
  \textit{et~al.}}{2021}]{Nagaraju2021ApOpt..60.8145N}
\begin{barticle}
\bauthor{\bsnm{{Nagaraju}}, \binits{K.}},
\bauthor{\bsnm{{Prasad}}, \binits{B.R.}},
\bauthor{\bsnm{{Hegde}}, \binits{B.S.}},
\bauthor{\bsnm{{Narra}}, \binits{S.V.}},
\bauthor{\bsnm{{Utkarsha}}, \binits{D.}},
\bauthor{\bsnm{{Kumar}}, \binits{A.}},
\bauthor{\bsnm{{Singh}}, \binits{J.}},
\bauthor{\bsnm{{Kumar}}, \binits{V.}}:
\byear{2021},
\batitle{{Spectropolarimeter on board the Aditya-L1: polarization modulation
  and demodulation}}.
\bjtitle{Applied Optics}
\bvolume{60}(\bissue{26}),
\bfpage{8145}.
\doiurl{10.1364/AO.434219}.
\adsurl{2021ApOpt..60.8145N}.
\end{barticle}
\endbibitem

\bibitem[\protect\citeauthoryear{{Peter}
  \textit{et~al.}}{2012}]{Peter2012ExA....33..271P}
\begin{barticle}
\bauthor{\bsnm{{Peter}}, \binits{H.}},
\bauthor{\bsnm{{Abbo}}, \binits{L.}},
\bauthor{\bsnm{{Andretta}}, \binits{V.}},
\bauthor{\bsnm{{Auch{\`e}re}}, \binits{F.}},
\bauthor{\bsnm{{Bemporad}}, \binits{A.}},
\bauthor{\bsnm{{Berrilli}}, \binits{F.}},
\bauthor{\bsnm{{Bommier}}, \binits{V.}},
\bauthor{\bsnm{{Braukhane}}, \binits{A.}},
\bauthor{\bsnm{{Casini}}, \binits{R.}},
\bauthor{\bsnm{{Curdt}}, \binits{W.}},
\bauthor{\bsnm{{Davila}}, \binits{J.}},
\bauthor{\bsnm{{Dittus}}, \binits{H.}},
\bauthor{\bsnm{{Fineschi}}, \binits{S.}},
\bauthor{\bsnm{{Fludra}}, \binits{A.}},
\bauthor{\bsnm{{Gandorfer}}, \binits{A.}},
\bauthor{\bsnm{{Griffin}}, \binits{D.}},
\bauthor{\bsnm{{Inhester}}, \binits{B.}},
\bauthor{\bsnm{{Lagg}}, \binits{A.}},
\bauthor{\bsnm{{Landi Degl'Innocenti}}, \binits{E.}},
\bauthor{\bsnm{{Maiwald}}, \binits{V.}},
\bauthor{\bsnm{{Sainz}}, \binits{R.M.}},
\bauthor{\bsnm{{Mart{\'\i}nez Pillet}}, \binits{V.}},
\bauthor{\bsnm{{Matthews}}, \binits{S.}},
\bauthor{\bsnm{{Moses}}, \binits{D.}},
\bauthor{\bsnm{{Parenti}}, \binits{S.}},
\bauthor{\bsnm{{Pietarila}}, \binits{A.}},
\bauthor{\bsnm{{Quantius}}, \binits{D.}},
\bauthor{\bsnm{{Raouafi}}, \binits{N.-E.}},
\bauthor{\bsnm{{Raymond}}, \binits{J.}},
\bauthor{\bsnm{{Rochus}}, \binits{P.}},
\bauthor{\bsnm{{Romberg}}, \binits{O.}},
\bauthor{\bsnm{{Schlotterer}}, \binits{M.}},
\bauthor{\bsnm{{Sch{\"u}hle}}, \binits{U.}},
\bauthor{\bsnm{{Solanki}}, \binits{S.}},
\bauthor{\bsnm{{Spadaro}}, \binits{D.}},
\bauthor{\bsnm{{Teriaca}}, \binits{L.}},
\bauthor{\bsnm{{Tomczyk}}, \binits{S.}},
\bauthor{\bsnm{{Trujillo Bueno}}, \binits{J.}},
\bauthor{\bsnm{{Vial}}, \binits{J.-C.}}:
\byear{2012},
\batitle{{Solar magnetism eXplorer (SolmeX). Exploring the magnetic field in
  the upper atmosphere of our closest star}}.
\bjtitle{Experimental Astronomy}
\bvolume{33}(\bissue{2-3}),
\bfpage{271}.
\doiurl{10.1007/s10686-011-9271-0}.
\adsurl{2012ExA....33..271P}.
\end{barticle}
\endbibitem

\bibitem[\protect\citeauthoryear{{Priest} and
  {Hood}}{1991}]{Priest1991gamp.conf.....P}
\begin{bbook}
\bauthor{\bsnm{{Priest}}, \binits{E.R.}},
\bauthor{\bsnm{{Hood}}, \binits{A.W.}}:
\byear{1991},
\bbtitle{{Advances in solar system magnetohydrodynamics}}.
\adsurl{1991gamp.conf.....P}.
\end{bbook}
\endbibitem

\bibitem[\protect\citeauthoryear{{Querfeld} and
  {Smartt}}{1984}]{Querfeld1984SoPh...91..299Q}
\begin{barticle}
\bauthor{\bsnm{{Querfeld}}, \binits{C.W.}},
\bauthor{\bsnm{{Smartt}}, \binits{R.N.}}:
\byear{1984},
\batitle{{Comparison of Coronal Emission Line Structure and Polarization}}.
\bjtitle{\solphys}
\bvolume{91}(\bissue{2}),
\bfpage{299}.
\doiurl{10.1007/BF00146301}.
\adsurl{1984SoPh...91..299Q}.
\end{barticle}
\endbibitem

\bibitem[\protect\citeauthoryear{{Raghavendra Prasad}
  \textit{et~al.}}{2017}]{RaghavendraPrasad2017CSci..113..613R}
\begin{barticle}
\bauthor{\bsnm{{Raghavendra Prasad}}, \binits{B.}},
\bauthor{\bsnm{{Banerjee}}, \binits{D.}},
\bauthor{\bsnm{{Singh}}, \binits{J.}},
\bauthor{\bsnm{{Nagabhushana}}, \binits{S.}},
\bauthor{\bsnm{{Kumar}}, \binits{A.}},
\bauthor{\bsnm{{Kamath}}, \binits{P.U.}},
\bauthor{\bsnm{{Kathiravan}}, \binits{S.}},
\bauthor{\bsnm{{Venkata}}, \binits{S.}},
\bauthor{\bsnm{{Rajkumar}}, \binits{N.}},
\bauthor{\bsnm{{Natarajan}}, \binits{V.}},
\bauthor{\bsnm{{Juneja}}, \binits{M.}},
\bauthor{\bsnm{{Somu}}, \binits{P.}},
\bauthor{\bsnm{{Pant}}, \binits{V.}},
\bauthor{\bsnm{{Shaji}}, \binits{N.}},
\bauthor{\bsnm{{Sankarsubramanian}}, \binits{K.}},
\bauthor{\bsnm{{Patra}}, \binits{A.}},
\bauthor{\bsnm{{Venkateswaran}}, \binits{R.}},
\bauthor{\bsnm{{Adoni}}, \binits{A.A.}},
\bauthor{\bsnm{{Narendra}}, \binits{S.}},
\bauthor{\bsnm{{Haridas}}, \binits{T.R.}},
\bauthor{\bsnm{{Mathew}}, \binits{S.K.}},
\bauthor{\bsnm{{Mohan Krishna}}, \binits{R.}},
\bauthor{\bsnm{{Amareswari}}, \binits{K.}},
\bauthor{\bsnm{{Jaiswal}}, \binits{B.}}:
\byear{2017},
\batitle{{Visible Emission Line Coronagraph on Aditya-L1}}.
\bjtitle{Current Science}
\bvolume{113}(\bissue{4}),
\bfpage{613}.
\adsurl{2017CSci..113..613R}.
\end{barticle}
\endbibitem

\bibitem[\protect\citeauthoryear{{Raouafi}, {Lemaire}, and
  {Sahal-Br{\'e}chot}}{1999}]{RaouafiOVIdetect1999A&A...345..999R}
\begin{barticle}
\bauthor{\bsnm{{Raouafi}}, \binits{N.-E.}},
\bauthor{\bsnm{{Lemaire}}, \binits{P.}},
\bauthor{\bsnm{{Sahal-Br{\'e}chot}}, \binits{S.}}:
\byear{1999},
\batitle{{Detection of the O VI 103.2 NM line polarization by the SUMER
  spectrometer on the SOHO spacecraft}}.
\bjtitle{\aap}
\bvolume{345},
\bfpage{999}.
\adsurl{1999A&A...345..999R}.
\end{barticle}
\endbibitem

\bibitem[\protect\citeauthoryear{{Raouafi}, {Sahal-Br{\'e}chot}, and
  {Lemaire}}{2002}]{RaouafiOVIBmeasure2002A&A...396.1019R}
\begin{barticle}
\bauthor{\bsnm{{Raouafi}}, \binits{N.-E.}},
\bauthor{\bsnm{{Sahal-Br{\'e}chot}}, \binits{S.}},
\bauthor{\bsnm{{Lemaire}}, \binits{P.}}:
\byear{2002},
\batitle{{Linear polarization of the O VI lambda 1031.92 coronal line. II.
  Constraints on the magnetic field and the solar wind velocity field vectors
  in the coronal polar holes}}.
\bjtitle{\aap}
\bvolume{396},
\bfpage{1019}.
\doiurl{10.1051/0004-6361:20021418}.
\adsurl{2002A&A...396.1019R}.
\end{barticle}
\endbibitem

\bibitem[\protect\citeauthoryear{{Raouafi}
  \textit{et~al.}}{2002}]{RaouafiOVImeasurepol2002A&A...390..691R}
\begin{barticle}
\bauthor{\bsnm{{Raouafi}}, \binits{N.-E.}},
\bauthor{\bsnm{{Sahal-Br{\'e}chot}}, \binits{S.}},
\bauthor{\bsnm{{Lemaire}}, \binits{P.}},
\bauthor{\bsnm{{Bommier}}, \binits{V.}}:
\byear{2002},
\batitle{{Linear polarization of the O VI lambda 1031.92 coronal line. I.
  Constraints on the solar wind velocity field vector in the polar holes}}.
\bjtitle{\aap}
\bvolume{390},
\bfpage{691}.
\doiurl{10.1051/0004-6361:20020752}.
\adsurl{2002A&A...390..691R}.
\end{barticle}
\endbibitem

\bibitem[\protect\citeauthoryear{{Rast}
  \textit{et~al.}}{2021}]{Rast2021SoPh..296...70R}
\begin{barticle}
\bauthor{\bsnm{{Rast}}, \binits{M.P.}},
\bauthor{\bsnm{{Bello Gonz{\'a}lez}}, \binits{N.}},
\bauthor{\bsnm{{Bellot Rubio}}, \binits{L.}},
\bauthor{\bsnm{{Cao}}, \binits{W.}},
\bauthor{\bsnm{{Cauzzi}}, \binits{G.}},
\bauthor{\bsnm{{Deluca}}, \binits{E.}},
\bauthor{\bsnm{{de Pontieu}}, \binits{B.}},
\bauthor{\bsnm{{Fletcher}}, \binits{L.}},
\bauthor{\bsnm{{Gibson}}, \binits{S.E.}},
\bauthor{\bsnm{{Judge}}, \binits{P.G.}},
\bauthor{\bsnm{{Katsukawa}}, \binits{Y.}},
\bauthor{\bsnm{{Kazachenko}}, \binits{M.D.}},
\bauthor{\bsnm{{Khomenko}}, \binits{E.}},
\bauthor{\bsnm{{Landi}}, \binits{E.}},
\bauthor{\bsnm{{Mart{\'\i}nez Pillet}}, \binits{V.}},
\bauthor{\bsnm{{Petrie}}, \binits{G.J.D.}},
\bauthor{\bsnm{{Qiu}}, \binits{J.}},
\bauthor{\bsnm{{Rachmeler}}, \binits{L.A.}},
\bauthor{\bsnm{{Rempel}}, \binits{M.}},
\bauthor{\bsnm{{Schmidt}}, \binits{W.}},
\bauthor{\bsnm{{Scullion}}, \binits{E.}},
\bauthor{\bsnm{{Sun}}, \binits{X.}},
\bauthor{\bsnm{{Welsch}}, \binits{B.T.}},
\bauthor{\bsnm{{Andretta}}, \binits{V.}},
\bauthor{\bsnm{{Antolin}}, \binits{P.}},
\bauthor{\bsnm{{Ayres}}, \binits{T.R.}},
\bauthor{\bsnm{{Balasubramaniam}}, \binits{K.S.}},
\bauthor{\bsnm{{Ballai}}, \binits{I.}},
\bauthor{\bsnm{{Berger}}, \binits{T.E.}},
\bauthor{\bsnm{{Bradshaw}}, \binits{S.J.}},
\bauthor{\bsnm{{Campbell}}, \binits{R.J.}},
\bauthor{\bsnm{{Carlsson}}, \binits{M.}},
\bauthor{\bsnm{{Casini}}, \binits{R.}},
\bauthor{\bsnm{{Centeno}}, \binits{R.}},
\bauthor{\bsnm{{Cranmer}}, \binits{S.R.}},
\bauthor{\bsnm{{Criscuoli}}, \binits{S.}},
\bauthor{\bsnm{{Deforest}}, \binits{C.}},
\bauthor{\bsnm{{Deng}}, \binits{Y.}},
\bauthor{\bsnm{{Erd{\'e}lyi}}, \binits{R.}},
\bauthor{\bsnm{{Fedun}}, \binits{V.}},
\bauthor{\bsnm{{Fischer}}, \binits{C.E.}},
\bauthor{\bsnm{{Gonz{\'a}lez Manrique}}, \binits{S.J.}},
\bauthor{\bsnm{{Hahn}}, \binits{M.}},
\bauthor{\bsnm{{Harra}}, \binits{L.}},
\bauthor{\bsnm{{Henriques}}, \binits{V.M.J.}},
\bauthor{\bsnm{{Hurlburt}}, \binits{N.E.}},
\bauthor{\bsnm{{Jaeggli}}, \binits{S.}},
\bauthor{\bsnm{{Jafarzadeh}}, \binits{S.}},
\bauthor{\bsnm{{Jain}}, \binits{R.}},
\bauthor{\bsnm{{Jefferies}}, \binits{S.M.}},
\bauthor{\bsnm{{Keys}}, \binits{P.H.}},
\bauthor{\bsnm{{Kowalski}}, \binits{A.F.}},
\bauthor{\bsnm{{Kuckein}}, \binits{C.}},
\bauthor{\bsnm{{Kuhn}}, \binits{J.R.}},
\bauthor{\bsnm{{Kuridze}}, \binits{D.}},
\bauthor{\bsnm{{Liu}}, \binits{J.}},
\bauthor{\bsnm{{Liu}}, \binits{W.}},
\bauthor{\bsnm{{Longcope}}, \binits{D.}},
\bauthor{\bsnm{{Mathioudakis}}, \binits{M.}},
\bauthor{\bsnm{{McAteer}}, \binits{R.T.J.}},
\bauthor{\bsnm{{McIntosh}}, \binits{S.W.}},
\bauthor{\bsnm{{McKenzie}}, \binits{D.E.}},
\bauthor{\bsnm{{Miralles}}, \binits{M.P.}},
\bauthor{\bsnm{{Morton}}, \binits{R.J.}},
\bauthor{\bsnm{{Muglach}}, \binits{K.}},
\bauthor{\bsnm{{Nelson}}, \binits{C.J.}},
\bauthor{\bsnm{{Panesar}}, \binits{N.K.}},
\bauthor{\bsnm{{Parenti}}, \binits{S.}},
\bauthor{\bsnm{{Parnell}}, \binits{C.E.}},
\bauthor{\bsnm{{Poduval}}, \binits{B.}},
\bauthor{\bsnm{{Reardon}}, \binits{K.P.}},
\bauthor{\bsnm{{Reep}}, \binits{J.W.}},
\bauthor{\bsnm{{Schad}}, \binits{T.A.}},
\bauthor{\bsnm{{Schmit}}, \binits{D.}},
\bauthor{\bsnm{{Sharma}}, \binits{R.}},
\bauthor{\bsnm{{Socas-Navarro}}, \binits{H.}},
\bauthor{\bsnm{{Srivastava}}, \binits{A.K.}},
\bauthor{\bsnm{{Sterling}}, \binits{A.C.}},
\bauthor{\bsnm{{Suematsu}}, \binits{Y.}},
\bauthor{\bsnm{{Tarr}}, \binits{L.A.}},
\bauthor{\bsnm{{Tiwari}}, \binits{S.}},
\bauthor{\bsnm{{Tritschler}}, \binits{A.}},
\bauthor{\bsnm{{Verth}}, \binits{G.}},
\bauthor{\bsnm{{Vourlidas}}, \binits{A.}},
\bauthor{\bsnm{{Wang}}, \binits{H.}},
\bauthor{\bsnm{{Wang}}, \binits{Y.-M.}},
\bauthor{\bsnm{{NSO and DKIST Project}}},
\bauthor{\bsnm{{DKIST Instrument Scientists}}},
\bauthor{\bsnm{{DKIST Science Working Group}}},
\bauthor{\bsnm{{DKIST Critical Science Plan Community}}}:
\byear{2021},
\batitle{{Critical Science Plan for the Daniel K. Inouye Solar Telescope
  (DKIST)}}.
\bjtitle{\solphys}
\bvolume{296}(\bissue{4}),
\bfpage{70}.
\doiurl{10.1007/s11207-021-01789-2}.
\adsurl{2021SoPh..296...70R}.
\end{barticle}
\endbibitem

\bibitem[\protect\citeauthoryear{{Sahal-Brechot}}{1977}]{Sahal1977ApJ...213..887S}
\begin{barticle}
\bauthor{\bsnm{{Sahal-Brechot}}, \binits{S.}}:
\byear{1977},
\batitle{{Calculation of the polarization degree of the infrared lines of Fe
  XIII of the solar corona.}}
\bjtitle{\apj}
\bvolume{213},
\bfpage{887}.
\doiurl{10.1086/155221}.
\adsurl{1977ApJ...213..887S}.
\end{barticle}
\endbibitem

\bibitem[\protect\citeauthoryear{{Sahal-Brechot}}{1981}]{Brechot1981SSRv...29..391S}
\begin{barticle}
\bauthor{\bsnm{{Sahal-Brechot}}, \binits{S.}}:
\byear{1981},
\batitle{{The Hanle effect applied to magnetic field diagnostics.}}
\bjtitle{\ssr}
\bvolume{29},
\bfpage{391}.
\adsurl{1981SSRv...29..391S}.
\end{barticle}
\endbibitem

\bibitem[\protect\citeauthoryear{{Sahal-Brechot}, {Malinovsky}, and
  {Bommier}}{1986}]{SahalOVI1986A&A...168..284S}
\begin{barticle}
\bauthor{\bsnm{{Sahal-Brechot}}, \binits{S.}},
\bauthor{\bsnm{{Malinovsky}}, \binits{M.}},
\bauthor{\bsnm{{Bommier}}, \binits{V.}}:
\byear{1986},
\batitle{{The polarization of the O VI 1032 {\r{A}} line as a probe for
  measuringthe coronal vector magnetic field via the Hanle effect.}}
\bjtitle{\aap}
\bvolume{168},
\bfpage{284}.
\adsurl{1986A&A...168..284S}.
\end{barticle}
\endbibitem

\bibitem[\protect\citeauthoryear{{Saqri}
  \textit{et~al.}}{2020}]{Saqri2020SoPh..295....6S}
\begin{barticle}
\bauthor{\bsnm{{Saqri}}, \binits{J.}},
\bauthor{\bsnm{{Veronig}}, \binits{A.M.}},
\bauthor{\bsnm{{Heinemann}}, \binits{S.G.}},
\bauthor{\bsnm{{Hofmeister}}, \binits{S.J.}},
\bauthor{\bsnm{{Temmer}}, \binits{M.}},
\bauthor{\bsnm{{Dissauer}}, \binits{K.}},
\bauthor{\bsnm{{Su}}, \binits{Y.}}:
\byear{2020},
\batitle{{Differential Emission Measure Plasma Diagnostics of a Long-Lived
  Coronal Hole}}.
\bjtitle{\solphys}
\bvolume{295}(\bissue{1}),
\bfpage{6}.
\doiurl{10.1007/s11207-019-1570-z}.
\adsurl{2020SoPh..295....6S}.
\end{barticle}
\endbibitem

\bibitem[\protect\citeauthoryear{{Sasikumar Raja}
  \textit{et~al.}}{2021}]{Sasikumar2021arXiv211014179S}
\begin{botherref}
\oauthor{\bsnm{{Sasikumar Raja}}, \binits{K.}},
\oauthor{\bsnm{{Venkata}}, \binits{S.}},
\oauthor{\bsnm{{Singh}}, \binits{J.}},
\oauthor{\bsnm{{Raghavendra Prasad}}, \binits{B.}}:
2021,
{Solar Coronal Magnetic Fields and Sensitivity Requirements for
  Spectropolarimetry Channel of VELC Onboard Aditya-L1}.
\textit{arXiv e-prints},
arXiv:2110.14179.
\adsurl{2021arXiv211014179S}.
\end{botherref}
\endbibitem

\bibitem[\protect\citeauthoryear{{Stenflo}}{1994}]{Stenflo1994ASSL..189.....S}
\begin{bbook}
\bauthor{\bsnm{{Stenflo}}, \binits{J.}}:
\byear{1994},
\bbtitle{{Solar Magnetic Fields: Polarized Radiation Diagnostics}}
\bseriesno{189}.
\doiurl{10.1007/978-94-015-8246-9}.
\adsurl{1994ASSL..189.....S}.
\end{bbook}
\endbibitem

\bibitem[\protect\citeauthoryear{{Tian}
  \textit{et~al.}}{2009}]{Tian2009A&A...504..239T}
\begin{barticle}
\bauthor{\bsnm{{Tian}}, \binits{H.}},
\bauthor{\bsnm{{Curdt}}, \binits{W.}},
\bauthor{\bsnm{{Marsch}}, \binits{E.}},
\bauthor{\bsnm{{Sch{\"u}hle}}, \binits{U.}}:
\byear{2009},
\batitle{{Hydrogen Lyman-{\ensuremath{\alpha}} and Lyman-{\ensuremath{\beta}}
  spectral radiance profiles in the quiet Sun}}.
\bjtitle{\aap}
\bvolume{504}(\bissue{1}),
\bfpage{239}.
\doiurl{10.1051/0004-6361/200811445}.
\adsurl{2009A&A...504..239T}.
\end{barticle}
\endbibitem

\bibitem[\protect\citeauthoryear{{Tomczyk}
  \textit{et~al.}}{2008}]{Tomczyk2008SoPh..247..411T}
\begin{barticle}
\bauthor{\bsnm{{Tomczyk}}, \binits{S.}},
\bauthor{\bsnm{{Card}}, \binits{G.L.}},
\bauthor{\bsnm{{Darnell}}, \binits{T.}},
\bauthor{\bsnm{{Elmore}}, \binits{D.F.}},
\bauthor{\bsnm{{Lull}}, \binits{R.}},
\bauthor{\bsnm{{Nelson}}, \binits{P.G.}},
\bauthor{\bsnm{{Streander}}, \binits{K.V.}},
\bauthor{\bsnm{{Burkepile}}, \binits{J.}},
\bauthor{\bsnm{{Casini}}, \binits{R.}},
\bauthor{\bsnm{{Judge}}, \binits{P.G.}}:
\byear{2008},
\batitle{{An Instrument to Measure Coronal Emission Line Polarization}}.
\bjtitle{\solphys}
\bvolume{247}(\bissue{2}),
\bfpage{411}.
\doiurl{10.1007/s11207-007-9103-6}.
\adsurl{2008SoPh..247..411T}.
\end{barticle}
\endbibitem

\bibitem[\protect\citeauthoryear{{Tomczyk}
  \textit{et~al.}}{2016}]{Tomczyk2016JGRA..121.7470T}
\begin{barticle}
\bauthor{\bsnm{{Tomczyk}}, \binits{S.}},
\bauthor{\bsnm{{Landi}}, \binits{E.}},
\bauthor{\bsnm{{Burkepile}}, \binits{J.T.}},
\bauthor{\bsnm{{Casini}}, \binits{R.}},
\bauthor{\bsnm{{DeLuca}}, \binits{E.E.}},
\bauthor{\bsnm{{Fan}}, \binits{Y.}},
\bauthor{\bsnm{{Gibson}}, \binits{S.E.}},
\bauthor{\bsnm{{Lin}}, \binits{H.}},
\bauthor{\bsnm{{McIntosh}}, \binits{S.W.}},
\bauthor{\bsnm{{Solomon}}, \binits{S.C.}},
\bauthor{\bsnm{{Toma}}, \binits{G.}},
\bauthor{\bsnm{{Wijn}}, \binits{A.G.}},
\bauthor{\bsnm{{Zhang}}, \binits{J.}}:
\byear{2016},
\batitle{{Scientific objectives and capabilities of the Coronal Solar Magnetism
  Observatory}}.
\bjtitle{Journal of Geophysical Research (Space Physics)}
\bvolume{121}(\bissue{8}),
\bfpage{7470}.
\doiurl{10.1002/2016JA022871}.
\adsurl{2016JGRA..121.7470T}.
\end{barticle}
\endbibitem

\bibitem[\protect\citeauthoryear{{Trujillo
  Bueno}}{2014}]{Trujillo2014ASPC..489..137T}
\begin{bchapter}
\bauthor{\bsnm{{Trujillo Bueno}}, \binits{J.}}:
\byear{2014},
\bctitle{{Polarized Radiation Observables for Probing the Magnetism of the
  Outer Solar Atmosphere}}.
In: \beditor{\bsnm{{Nagendra}}, \binits{K.N.}},
\beditor{\bsnm{{Stenflo}}, \binits{J.O.}},
\beditor{\bsnm{{Qu}}, \binits{Z.Q.}},
\beditor{\bsnm{{Sampoorna}}, \binits{M.}} (eds.)
\bbtitle{Solar Polarization 7},
\bsertitle{Astronomical Society of the Pacific Conference Series}
\bseriesno{489},
\bfpage{137}.
\adsurl{2014ASPC..489..137T}.
\end{bchapter}
\endbibitem

\bibitem[\protect\citeauthoryear{{Trujillo Bueno}, {Landi Degl'Innocenti}, and
  {Belluzzi}}{2017}]{Trujillo2017SSRv..210..183T}
\begin{barticle}
\bauthor{\bsnm{{Trujillo Bueno}}, \binits{J.}},
\bauthor{\bsnm{{Landi Degl'Innocenti}}, \binits{E.}},
\bauthor{\bsnm{{Belluzzi}}, \binits{L.}}:
\byear{2017},
\batitle{{The Physics and Diagnostic Potential of Ultraviolet
  Spectropolarimetry}}.
\bjtitle{\ssr}
\bvolume{210}(\bissue{1-4}),
\bfpage{183}.
\doiurl{10.1007/s11214-016-0306-8}.
\adsurl{2017SSRv..210..183T}.
\end{barticle}
\endbibitem

\bibitem[\protect\citeauthoryear{{Verdebout}
  \textit{et~al.}}{2014}]{verdebout2014}
\begin{barticle}
\bauthor{\bsnm{{Verdebout}}, \binits{S.}},
\bauthor{\bsnm{{Naz{\'e}}}, \binits{C.}},
\bauthor{\bsnm{{J{\"o}nsson}}, \binits{P.}},
\bauthor{\bsnm{{Rynkun}}, \binits{P.}},
\bauthor{\bsnm{{Godefroid}}, \binits{M.}},
\bauthor{\bsnm{{Gaigalas}}, \binits{G.}}:
\byear{2014},
\batitle{{Hyperfine structures and Land{\'e} g$_{J}$-factors for n=2 states in
  beryllium-, boron-, carbon-, and nitrogen-like ions from relativistic
  configuration interaction calculations}}.
\bjtitle{Atomic Data and Nuclear Data Tables}
\bvolume{100}(\bissue{5}),
\bfpage{1111}.
\doiurl{10.1016/j.adt.2014.05.001}.
\adsurl{2014ADNDT.100.1111V}.
\end{barticle}
\endbibitem

\bibitem[\protect\citeauthoryear{{Warren} and
  {Brooks}}{2009}]{Warren2009ApJ...700..762W}
\begin{barticle}
\bauthor{\bsnm{{Warren}}, \binits{H.P.}},
\bauthor{\bsnm{{Brooks}}, \binits{D.H.}}:
\byear{2009},
\batitle{{The Temperature and Density Structure of the Solar Corona. I.
  Observations of the Quiet Sun with the EUV Imaging Spectrometer on Hinode}}.
\bjtitle{\apj}
\bvolume{700}(\bissue{1}),
\bfpage{762}.
\doiurl{10.1088/0004-637X/700/1/762}.
\adsurl{2009ApJ...700..762W}.
\end{barticle}
\endbibitem

\bibitem[\protect\citeauthoryear{{Warren} and
  {Warshall}}{2002}]{Warren2002ApJ...571..999W}
\begin{barticle}
\bauthor{\bsnm{{Warren}}, \binits{H.P.}},
\bauthor{\bsnm{{Warshall}}, \binits{A.D.}}:
\byear{2002},
\batitle{{Temperature and Density Measurements in a Quiet Coronal Streamer}}.
\bjtitle{\apj}
\bvolume{571}(\bissue{2}),
\bfpage{999}.
\doiurl{10.1086/340069}.
\adsurl{2002ApJ...571..999W}.
\end{barticle}
\endbibitem

\bibitem[\protect\citeauthoryear{{Yang}
  \textit{et~al.}}{2020}]{YangGlobal2020Sci...369..694Y}
\begin{barticle}
\bauthor{\bsnm{{Yang}}, \binits{Z.}},
\bauthor{\bsnm{{Bethge}}, \binits{C.}},
\bauthor{\bsnm{{Tian}}, \binits{H.}},
\bauthor{\bsnm{{Tomczyk}}, \binits{S.}},
\bauthor{\bsnm{{Morton}}, \binits{R.}},
\bauthor{\bsnm{{Del Zanna}}, \binits{G.}},
\bauthor{\bsnm{{McIntosh}}, \binits{S.W.}},
\bauthor{\bsnm{{Karak}}, \binits{B.B.}},
\bauthor{\bsnm{{Gibson}}, \binits{S.}},
\bauthor{\bsnm{{Samanta}}, \binits{T.}},
\bauthor{\bsnm{{He}}, \binits{J.}},
\bauthor{\bsnm{{Chen}}, \binits{Y.}},
\bauthor{\bsnm{{Wang}}, \binits{L.}}:
\byear{2020},
\batitle{{Global maps of the magnetic field in the solar corona}}.
\bjtitle{Science}
\bvolume{369}(\bissue{6504}),
\bfpage{694}.
\doiurl{10.1126/science.abb4462}.
\adsurl{2020Sci...369..694Y}.
\end{barticle}
\endbibitem

\bibitem[\protect\citeauthoryear{{Young} \textit{et~al.}}{2007}]{young2007euv}
\begin{barticle}
\bauthor{\bsnm{{Young}}, \binits{P.R.}},
\bauthor{\bsnm{{Del Zanna}}, \binits{G.}},
\bauthor{\bsnm{{Mason}}, \binits{H.E.}},
\bauthor{\bsnm{{Dere}}, \binits{K.P.}},
\bauthor{\bsnm{{Landi}}, \binits{E.}},
\bauthor{\bsnm{{Landini}}, \binits{M.}},
\bauthor{\bsnm{{Doschek}}, \binits{G.A.}},
\bauthor{\bsnm{{Brown}}, \binits{C.M.}},
\bauthor{\bsnm{{Culhane}}, \binits{L.}},
\bauthor{\bsnm{{Harra}}, \binits{L.K.}},
\bauthor{\bsnm{{Watanabe}}, \binits{T.}},
\bauthor{\bsnm{{Hara}}, \binits{H.}}:
\byear{2007},
\batitle{{EUV Emission Lines and Diagnostics Observed with Hinode/EIS}}.
\bjtitle{\pasj}
\bvolume{59},
\bfpage{S857}.
\doiurl{10.1093/pasj/59.sp3.S857}.
\adsurl{2007PASJ...59S.857Y}.
\end{barticle}
\endbibitem

\bibitem[\protect\citeauthoryear{{Zanna} and {Mason}}{2005}]{del2005spectral}
\begin{barticle}
\bauthor{\bsnm{{Zanna}}, \binits{G.D.}},
\bauthor{\bsnm{{Mason}}, \binits{H.E.}}:
\byear{2005},
\batitle{{Spectral diagnostic capabilities of Solar-B EIS}}.
\bjtitle{Advances in Space Research}
\bvolume{36}(\bissue{8}),
\bfpage{1503}.
\doiurl{10.1016/j.asr.2005.03.122}.
\adsurl{2005AdSpR..36.1503Z}.
\end{barticle}
\endbibitem

\bibitem[\protect\citeauthoryear{{Zhao}
  \textit{et~al.}}{2019}]{Zhao2019ApJ...883...55Z}
\begin{barticle}
\bauthor{\bsnm{{Zhao}}, \binits{J.}},
\bauthor{\bsnm{{Gibson}}, \binits{S.E.}},
\bauthor{\bsnm{{Fineschi}}, \binits{S.}},
\bauthor{\bsnm{{Susino}}, \binits{R.}},
\bauthor{\bsnm{{Casini}}, \binits{R.}},
\bauthor{\bsnm{{Li}}, \binits{H.}},
\bauthor{\bsnm{{Gan}}, \binits{W.}}:
\byear{2019},
\batitle{{Simulating the Solar Corona in the Forbidden and Permitted Lines with
  Forward Modeling. I. Saturated and Unsaturated Hanle Regimes}}.
\bjtitle{\apj}
\bvolume{883}(\bissue{1}),
\bfpage{55}.
\doiurl{10.3847/1538-4357/ab328b}.
\adsurl{2019ApJ...883...55Z}.
\end{barticle}
\endbibitem

\bibitem[\protect\citeauthoryear{{Zhao}
  \textit{et~al.}}{2021}]{Zhao2021ApJ...912..141Z}
\begin{barticle}
\bauthor{\bsnm{{Zhao}}, \binits{J.}},
\bauthor{\bsnm{{Gibson}}, \binits{S.E.}},
\bauthor{\bsnm{{Fineschi}}, \binits{S.}},
\bauthor{\bsnm{{Susino}}, \binits{R.}},
\bauthor{\bsnm{{Casini}}, \binits{R.}},
\bauthor{\bsnm{{Cranmer}}, \binits{S.R.}},
\bauthor{\bsnm{{Ofman}}, \binits{L.}},
\bauthor{\bsnm{{Li}}, \binits{H.}}:
\byear{2021},
\batitle{{Simulating the Solar Minimum Corona in UV Wavelengths with Forward
  Modeling II. Doppler Dimming and Microscopic Anisotropy Effect}}.
\bjtitle{\apj}
\bvolume{912}(\bissue{2}),
\bfpage{141}.
\doiurl{10.3847/1538-4357/abf143}.
\adsurl{2021ApJ...912..141Z}.
\end{barticle}
\endbibitem

\end{thebibliography}

\end{article} 

\end{document}